\documentstyle[preprint,aps,eqsecnum]{revtex}
\begin{document}
\draft
\title{Exact solution of the Landau fixed point via bosonization}
\author{A.~H.~Castro Neto and Eduardo H.~Fradkin}
\bigskip
\address
{Loomis Laboratory of Physics\\
University of Illinois at Urbana-Champaign\\
1110 W.Green St., Urbana, IL, 61801-3080}

\maketitle

\begin{abstract}
We study, via bosonization, the Landau fixed point for the
problem of interacting spinless fermions near the Fermi surface
in dimensions higher than one.
We rederive the bosonic representation of the Fermi operator and use it
to find the general form of the fermion propagator for the Landau fixed
point. Using a generalized Bogoliubov transformation we diagonalize
exactly the bosonized hamiltonian for the fixed point
and calculate the Fermion propagator (and the quasiparticle residue)
for isotropic interactions (independently of their strength).
We reexamine two well known problems in this context: the screening
of long range potentials and the Landau damping of gauge fields.
We also discuss the origin of the Luttinger fixed point in one
dimension in contrast with the Landau fixed point in higher dimensions.
\end{abstract}

\bigskip


\narrowtext

\section{Introduction}

During the last fifty years the Landau theory
has been a paradigm used to explain the experimental
behavior and electronic properties of quantum Fermi liquids \cite{Baym,pines}.
Initially the theory appeared as a phenomenological framework with a few
parameters fixed by experiments. The presence of
unknown parameters reflected, at that time, the lack
of a microscopic theory.
However, it was an extraordinary and necessary
first step. Landau himself also established the route for
the microscopic explanation for the validity of the theory. The Landau
theory became the main tool for the study of the effects of correlations
in electronic systems and its foundation was eventually established on
microscopic grounds using field theoretic methods
\cite{abrikosov,nozieres,kadanoff}.

Although the main idea behind the Fermi liquid theory is
quite simple, that is, the idea of a quasiparticle, its realization
in terms of microscopic calculations is far from that.
The idea is that when an electron interacts with other electrons it polarizes
its vicinity and a cloud is formed around it
(in a electronic system a hole is formed around the electron due
to the electron-electron repulsion \cite{pines}). In its motion the electron
carries an extra inertia due to the existence of the other
electrons. In more usual words, the quasiparticle is a
dressed electron. In this picture the interaction between the
electrons does not affect the intrinsic properties of the electron
(that is, its quantum numbers) but only its dynamics. In field theoretic
terms, the existence of a quasiparticle is related
to the presence of an isolated singularity in the one-particle
Green's function \cite{abrikosov}. This singularity produces a
Dirac delta peak in the spectral function of the Green's function.
In the non-interacting case all the weight of the spectral function
(or the quasiparticle residue, $Z_F$) is in the peak ($Z_F = 1$).
The ground state in this
case is a filled Fermi sea (due to Pauli's exclusion principle)
with a sharp singularity at the Fermi momentum which defines
the Fermi surface. When the interaction
is turned on the strength of the peak is weakened ($0 <Z_F < 1$)
but it is still infinitely sharp.
The rest of the spectral weight is carried by an incoherent background
which plays no essential role in Fermi liquid theory. The most
important consequence of the presence of this sharp peak is that the
ground state is still a filled Fermi sea with a weakened singularity.
The existence of quasiparticles
is thus directly related with the existence of a well defined Fermi
surface.

Fermi liquid theory has also been used as the starting point
in order to understand the nature of condensed states of matter
such as the superconducting state. The theory
which explains the behavior of superconductivity of simple
metals, the BCS theory, starts from the fact that the
normal state is a quantum liquid described
by the Landau theory \cite{bcs}. The properties of the superconducting
phase thus depend on the properties of the normal phase of the material.

The observed unusual properties of the normal phase
of the cuprates\cite{htc,varma} posed the question of the existence of new
states of condensed matter which are not described by the Landau theory of
Fermi liquids. In particular, the phenomenology of the normal states appears to
indicate the vanishing of the singularity at the Fermi surface ($Z_F =0$)
\cite{varma}. Many theoretical scenarios for such a non-Fermi liquid
behavior have been proposed \cite{anderson,urbana}.

The apparent failure of the conventional Landau theory of the Fermi
liquid in the context of the cuprates (and, perhaps, in more general
strongly correlated systems) has motivated a wide search for
alternative theoretical tools that, in principle, could handle the
effects of strong correlations. In one space dimension, where the
interactions are always strong as a result of the kinematical
constraints, bosonization\cite{bosonization} has emerged as the main
theoretical tool. Recently the problem of bosonizing a dense Fermi
system in arbitrary dimensions has been the focus of intense research.
The main ideas were introduced originally by Luther \cite{luther},
rediscovered recently by Haldane \cite{haldane} and developed in great
detail by Houghton and Marston \cite{houghton} and by us \cite{us}.
Here we will use the method of bosonization by coherent states
that we have developed recently\cite{us}.

In this paper we investigate a class of fixed point Hamiltonians
for interacting fermions which exhibit Landau-type behavior. The main
motivation of this work is to see how does the conventional behavior
predicted by Fermi liquid theory arises in the bosonization approach.
It is important to reexamine this well known problem not just as a check on
our methods but also since, unlike the conventional many-body perturbation
theory approach to Fermi liquid theory, bosonization
is not based on self consistent resummations of perturbation theory.
In principle, it should yield exact results for the low energy
behavior of the system. Thus, this is
a necessary step if these methods are to be applied to more interesting
physical systems in which the Landau theory is believed to fail, such as
the problem of a dense system of fermions coupled to dynamical gauge
fields. Such systems are central for the understanding of some
of the most novel approaches to the problem of High Temperature
Superconductors \cite{lee} and to the compressible states of the Fractional
Quantum Hall Effect \cite{ana,halperin}.

In previous papers\cite{us} we have examined the Landau fixed point.
Here we show by an explicit calculation that systems of fermions
(at finite density, relativistic or not, continuum or lattice) which
interact via scalar potentials in the absence of nesting or gauge
fields belong to this universality class.
We investigate the properties of the operators that create the physical
low-energy states of the system described by Fermi liquid theory.
We construct a Hilbert space of states which represents the
physical states close to the Fermi energy. The fixed point Hamiltonians
contain only marginal operators acting on these states. We also
characterize the relevant operators at these fixed points which are
connected with low energy instabilities of the system. We will not
consider here the problem of spin and magnetic excitations.

We have shown in our earlier work that it is possible to
bosonize an interacting spinless fermionic liquid, at long wavelengths,
in terms of operators which create particle-hole pairs close to
the Fermi surface \cite{us}. Our results showed that
the bosons of the theory are sound waves which propagate on
the Fermi surface and have no resemblance to free non-relativistic bosons.
Essentially these bosons are topologically constrained
to the Fermi surface and therefore they
propagate in a non-flat metric. The dynamics of these bosons is related
to the elastic properties of the Fermi surface, that is,
the Fermi surface sustains a surface tension when the quasiparticle
residue is non-zero ($Z_F \neq 0$) \cite{us}. When the surface tension
vanishes ($Z_F = 0$) a phase transition occurs at the Fermi surface
and many properties of the system change abruptly. We further showed
explicitly that the bosonized theory yields the correct thermodynamic
properties of Fermi liquids.

In this paper we begin by reviewing in some detail the bosonization
procedure we have introduced before \cite{us}. We use this approach to
develop the generating functional for the bosonic fields in terms
of coherent states which are coherent superposition of particle-hole
pairs and represent the distortions of the Fermi surface.
This representation for the generating functional allows us to
discuss the form of the fermion operator in terms of the bosons.
Here the similarity with the bosonization in one dimensional systems becomes
immediately clear. We also show that bosonization and non-Fermi liquid
behavior (or rather, {\it non-Landau behavior})
are {\it not} one and the same thing. We argue that bosonization is a
more general concept and that the non-Fermi liquid behavior of
one-dimensional systems is a product of the smallness of available
phase space which enhances the interactions.
In one dimension the constraints imposed by the conservation
laws couple the oscillations
of the Fermi surface (in this case only two Fermi points) in a
manner which will be discussed below. In dimensions higher the one
the number of degrees of freedom (or Fermi points) is infinite and
conservation laws alone are not enough to drive the system
away from the Fermi liquid fixed point.

We rederive an explicit formula for the fermion operator written in terms
of bosons and we use it to obtain the
correct free fermion propagator in the limit of
long wavelengths. This calculation reveals many important aspects
of the bosonization procedure we have been developing and confirms,
once more, the usefulness of this method.

Since the boson operator is a product of two particle
operators in terms of the original fermions, we show that a two particle
interaction (which is written as product of four particle operators)
can be written only in four different forms which are bilinears in
bosonic operators. Assuming an isotropic interaction for the fermions
(that is, no dependence of the interaction on the position of the
Fermi surface) we {\it diagonalize exactly} the problem in the thermodynamic
limit. We calculate explicitly the fermion propagator and the quasiparticle
residue for any strength of the potential. We show that, for local
interactions,
in dimensions greater than one, Fermi liquid behavior is {\it always}
expected. Recently, a perturbative approach in terms of the bosonic operators
has been developed in order to show that bosonization is able to reproduce
the expected behavior of the self-energy for Fermi liquids in two dimensions
\cite{houghton2}. Moreover, our methods give us hints for the possibility
of breakdown of the Fermi liquid theory due to singular interactions at
the Fermi surface \cite{next}.

We reexamined the problem of {\it dynamical screening} of the
fermion-fermion interactions. Unlike the conventional RPA approach, we
do not resum the bubble diagrams self consistently since bosonization is
a non-perturbative approach. Hence, one has to work with the bare
interaction and let the dynamics of the system decide. Thus,
instead of assuming a ``screening
first" scenario, we approach the problem of dynamical screening by
calculating directly the full fermion one-particle Green's function.
We show that dynamical screening of long range interactions is
due to the excitation of particle-hole fluctuations with momentum
transfer tangent to the Fermi surface. We further show that
assuming screening first, as it is
usually done in the conventional approach to Fermi liquid theory leads
to physically incorrect results for one dimensional systems.

We also discuss the problem of screening of external probes in the
bosonic language.
We show that while external scalar potentials are always screened,
external gauge fields are
not screened but get Landau damping instead (this is the case of a
non-superconducting material). The reason is well known,
liquids do not screen transverse oscillations at low frequency.
We also show how bosonization
can explain the differences between the screening of
scalar (longitudinal) and vector (gauge) fields and in particular
we obtain expressions for the susceptibilities (response functions) in
both cases. We rederive the well known fact that the RPA result \cite{pines}
is exact in the limit of long wavelengths.

The paper is organized as follows: in section II we review our
bosonization procedure introducing some details which were not
discussed in our previous works; using the coherent states
defined by the boson annihilation operator we obtain the generating
functional for the bosonic fields in section III; in section IV,
we discuss the form of the fermionic operator in terms of the
bosons and its relationship with the coherent state path integral
developed in the previous section; with this machinery at hand,
in section V we obtain the non-interacting one-particle Green's function;
in section VI we study what kind of interactions the bosonic hamiltonian
can have if we start with a two-body interaction between the fermions,
we show that we have only four kinds of terms which are
possible in the bosonic language and we are able to classify
them; in section VII using a generalized Bogoliubov transformation
we diagonalize exactly the bosonized hamiltonian
for fermions interacting via isotropic interactions in dimensions higher
than one; in section VIII we calculate the one particle
propagator and the quasiparticle residue as a function of the potential
strength for Fermi liquids; in sections IX and X
we obtain the well known results for the response functions for
scalar and vector fields, respectively, and in section XI
we discuss the differences between one and higher dimensions in the
context of bosonization. Section XII contains our conclusions.

\section{Bosonization}

The bosonization of a fermionic system is based on the
algebra obeyed by the densities and currents \cite{us}. This algebra
is obtained in a restricted Hilbert space which contains
the states close to the Fermi surface. The relevant operator,
which generates all the states in this Hilbert space is
defined as,
\begin{equation}
n_{\vec{q}}(\vec{k}) = c^{\dag}_{\vec{k}-\frac{\vec{q}}{2}}
c_{\vec{k}+\frac{\vec{q}}{2}}.
\end{equation}

In general the commutation relations between the operators in (2.1)
are written in terms of an expansion of operators. However,
in the restricted Hilbert space, we expect that the substitution
of the commutation relations by their expectation value on the
state $|FS\rangle$ (representing the filled Fermi sea) will generate all the
relevant dynamics of the interacting fermionic system. The commutation
relation is obtained for the case of small momenta {\it normal} to
the Fermi surface (there is no restriction for momenta {\it tangent}
of the Fermi surface). We have shown that, in this Hilbert space, the
commutation relation between the operators in (2.1) is written as
\cite{us},
\begin{equation}
\left[n_{\vec{q}}(\vec{k}),n_{\vec{-q'}}(\vec{k'})\right]=
\delta_{\vec{k},\vec{k'}} \delta_{\vec{q},\vec{q'}} \vec{q}
\cdot \vec{v}_{\vec{k}} \delta(\mu-\epsilon_{\vec{k}}).
\end{equation}

We begin by defining a complete set of one particle states with spectrum
$\epsilon_{\vec{k}}$ which is
used to build the full Hilbert space. The velocity of the particles is
defined in the usual way,
\begin{equation}
\vec{v}_{\vec{k}} = \nabla \epsilon_{\vec{k}}.
\end{equation}
The Fermi surface is defined by the set of vectors $\{\vec{k}_F\}$
which obey the relation,
\begin{equation}
\mu=\epsilon_{\vec{k}_F}
\end{equation}
where $\mu$ is the chemical potential of the system.

Although the particle-hole operators (2.1) have almost bosonic character, the
operators do not annihilate the reference state $|FS\rangle$.
We need to normal order these operator relative to this state.
Also, canonical bosonic commutation relations are only obeyed by
suitably smeared operators at each Fermi point.
We define creation and annihilation operators,
\begin{equation}
a_{\vec{q}}(\vec{k}_F) = \sum_{\vec{k}} \Phi_{\Lambda}(|\vec{k}-\vec{k}_F|)
\left(n_{\vec{q}}(\vec{k})
\Theta(\vec{v}_{\vec{k}_F} \cdot \vec{q}) +
n_{-\vec{q}}(\vec{k})\Theta(-\vec{v}_{\vec{k}_F} \cdot \vec{q})
\right)
\end{equation}
and
\begin{equation}
a^{\dag}_{\vec{q}}(\vec{k}_F) =
\sum_{\vec{k}} \Phi_{\Lambda}(|\vec{k}-\vec{k}_F|)
\left(n_{-\vec{q}}(\vec{k})
\Theta(\vec{v}_{\vec{k}_F} \cdot \vec{q}) +
n_{\vec{q}}(\vec{k})\Theta(-\vec{v}_{\vec{k}_F} \cdot \vec{q})
\right)
\end{equation}
where $\Theta(x) = 1(-1) $ if $x>0(<0)$
and $\Phi_{\Lambda}(|\vec{k}-\vec{k}_F|)$ is a dimensionless
smearing function
which keeps the vectors $\vec{k}$ close to $\vec{k}_F$ that is,
\begin{equation}
Lim_{\Lambda \to 0} \Phi_{\Lambda}(|\vec{k}-\vec{k}_F|) =
\delta_{\vec{k},\vec{k}_F}
\end{equation}
where $\Lambda$ can be viewed as a cut-off in momentum space.
The idea is to construct spheres of radius $\Lambda$ which cover all the
states on the Fermi surface \cite{haldane}. We parametrize each one
of these spheres
by the Fermi momentum $\vec{k}_F$, with $\vec{q}$ at the center of the spheres
(alternatively, we could also have constructed pill boxes of height $\Lambda$
and length $D$ instead of spheres \cite{houghton},
the difference will be immaterial
since all the physical quantities will not depend on the way
we introduce the cut-off). This construction will give good
results whenever the states in the problem have momentum close to
$\vec{k}_F$ with fluctuations of order $\vec{q}$ such that
$q<<\Lambda<<k_F$.

It is straightforward to see that, by construction, we have,
\begin{equation}
a_{\vec{q}}(\vec{k}_F) \mid FS \rangle = 0.
\end{equation}
And the smeared operators $a_{\vec{q}}(\vec{k}_F)$ are found to obey the
commutation relations,
\begin{equation}
\left[a_{\vec{q}}(\vec{k}_F),a^{\dag}_{\vec{q'}}(\vec{k'}_F)\right]=
N_{\Lambda}(\vec{k}_F) V \mid \vec{q}.\vec{v}_{\vec{k}_F} \mid
\delta_{\vec{k}_F,\vec{k'}_F} \left(\delta_{\vec{q},\vec{q'}}
+\delta_{\vec{q},-\vec{q'}}\right),
\end{equation}
where $N_{\Lambda}(\vec{k}_F)$ is a local density of states defined
as follows ($V$ is the volume of the system),
\begin{equation}
N_{\Lambda}(\vec{k}_F)= \frac{1}{V} \sum_{\vec{k}}
|\Phi_{\Lambda}(|\vec{k}-\vec{k}_F|)|^2 \delta(\mu-\epsilon_{\vec{k}}).
\end{equation}
Eqs. (2.8) and (2.9) show that these operators have bosonic character and
generate the restricted Hilbert space of interest.

The local density of states is a measure of the number of states
per unit of energy per solid angle in the Fermi surface.
In general, in the absence of Van Hove singularities, the local density
of states is well behaved and independent of the cut-off. In this case
we substitute it by its natural average
which is the total density of states $N(0)$,
\begin{equation}
N(0) = \frac{1}{V} \sum_{\vec{k}} \delta(\mu-\epsilon_{\vec{k}})
\end{equation}
divided by the solid angle on the Fermi surface $S_d = \int d\Omega$.
Indeed, in our previous articles where we have studied the transport
and thermodynamic properties of Fermi liquids we have used
$N_{\Lambda}(\vec{k}_F) =N(0)/S_d$ \cite{us}.

Although we could work directly with the operators defined above
it is usual to rescale the operators in such way to absorb the
density of states in its definition. We set,
\begin{equation}
b_{\vec{q}}(\vec{k}_F) = \left(N_{\Lambda}(\vec{k}_F) V
|\vec{q}.\vec{v}_{\vec{k}_F}|\right)^{-1/2} a_{\vec{q}}(\vec{k}_F)
\end{equation}
which obey the usual bosonic algebra (independent of the cut-off),
\begin{equation}
\left[b_{\vec{q}}(\vec{k}_F),b^{\dag}_{\vec{q'}}(\vec{k'}_F)\right]=
\delta_{\vec{k}_F,\vec{k'}_F} \left(\delta_{\vec{q},\vec{q'}}
+\delta_{\vec{q},-\vec{q'}}\right).
\end{equation}
These relations between bosonic operators, which have particle-hole
character, will be the basis of our work.

\section{Coherent states and Generating Functional}

Although we have a closed algebra and a reference state to work out
the physics, we do not have yet a clear physical interpretation for
the bosonic operators. We know from eq.~(2.1) that they are related
to particle-hole excitations. However, in order to develop more physical
intuition about the excitations created by these operators, we use
the coherent states associated with them.

The coherent states are defined via a unitary operator,
\begin{equation}
U([\phi]) = \exp\left( - \sum_{\vec{k}_F,\vec{q}}
\frac{1}{N_{\Lambda}(\vec{k}_F)
V \vec{q} \cdot \vec{v}_{\vec{k}_F}} \phi_{\vec{q}}(\vec{k}_F)
n_{-\vec{q}}(\vec{k}_F)
\right)
\end{equation}
where $U^{-1} = U^{\dag}$ implies
$\phi_{-\vec{q}}(\vec{k}_F)=\phi^{*}_{\vec{q}}
(\vec{k}_F)$.
The operator $U$ is a {\it functional} of the fields
$\phi_{\vec{q}}(\vec{k}_F)$
which are defined on the Fermi surface.

We can also rewrite (3.1) as,
\begin{equation}
U([\phi]) = \exp\left( - \sum_{\vec{k}_F,\vec{q},\vec{q}
\cdot \vec{v}_{\vec{k}_F}>0}
\frac{1}{N_{\Lambda}(\vec{k}_F) V |\vec{q} \cdot \vec{v}_{\vec{k}_F}|}
\left(\phi_{\vec{q}}(\vec{k}_F) n_{-\vec{q}}(\vec{k}_F) -
\phi^{*}_{\vec{q}}(\vec{k}_F) n_{\vec{q}}(\vec{k}_F)\right)\right)
\end{equation}
and using the definitions (2.5) and (2.12) we have,
\begin{equation}
U([\phi]) = \exp\left( - \sum_{\vec{k}_F,\vec{q},\vec{q} \cdot
\vec{v}_{\vec{k}_F} >0}
\left(\frac{1}{N_{\Lambda}(\vec{k}_F) V |\vec{q} \cdot
\vec{v}_{\vec{k}_F}|}\right)^{1/2}
\left(\phi_{\vec{q}}(\vec{k}_F) b^{\dag}_{\vec{q}}(\vec{k}_F) -
\phi^{*}_{\vec{q}}(\vec{k}_F) b_{\vec{q}}(\vec{k}_F)
\right)\right).
\end{equation}

The coherent state is defined as the evolution of the reference state via
the operator $U$,
\begin{equation}
|[\phi]\rangle = U([\phi])|FS\rangle.
\end{equation}
It is easy to show that this state is an eigenstate of the destruction
operator,
\begin{equation}
b_{\vec{q}}(\vec{k}_F)|[\phi]\rangle = - \left(\frac{1}{N_{\Lambda}
(\vec{k}_F) V |\vec{q} \cdot \vec{v}_{\vec{k}_F}|}\right)^{1/2}
\phi_{\vec{q}}(\vec{k}_F)|[\phi]\rangle.
\end{equation}
This last result has a clear physical meaning. It means that the
coherent state represents a deformation of the Fermi surface at
the point $\vec{k}_F$ in the direction of $\vec{q}$ due to a coherent
(collective) superposition of particle-hole pairs. Therefore, the
bosonic field $\phi_{\vec{q}}(\vec{k}_F)$ is a measure of this deformation
at that point. These bosonic fields are topologically
constrained excitations which propagate on the Fermi surface.

As usual with coherent states, we can show that they
are not orthogonal to each other and that they are overcomplete.
The overcompleteness of these states \cite{klauder} physically means that the
bosons which propagate on the Fermi surface are wavepackets \cite{us}.
This is fully consistent with the standard picture of the Landau
theory.

The propagation of the many-body system in time, from time $0$ to
time $t$, can be obtained by a
calculation of the $S$-matrix with initial and final states of
the kind considered above. The $S$-matrix is defined in the
following way,
\begin{equation}
\langle [\varphi],t| [\tilde{\varphi}],0\rangle =
\langle [\varphi]| e^{-i H t} |[\tilde{\varphi}]\rangle
\end{equation}
where $H$ is the Hamiltonian of the system.

Using the closure relation for these states it is easy to show that
it can also be written in terms of path integrals,
\begin{eqnarray}
\langle [\varphi],t|[\tilde{\varphi}],0\rangle =
\exp\left\{ -\sum_{\vec{k}_F,\vec{k'}_F}
\sum_{\vec{q},\vec{q} \cdot \vec{v}_{\vec{k}_F}>0}
\frac{1}{2 N_{\Lambda}(\vec{k}_F) V}
\left(\frac{1}{|\vec{q} \cdot \vec{v}_{\vec{k}_F}|}|
\varphi_{\vec{q}}(\vec{k}_F)|^2 +
\frac{1}{|\vec{q}.\vec{v}_{\vec{k'}_F}|}
\tilde{\varphi}_{\vec{q}}(\vec{k'}_F)|^2\right)\right\}
\nonumber
\\
\int^{[\varphi]}_{[\tilde{\varphi}]} D^2[\phi]
e^{i S[\phi]}
\end{eqnarray}
where $D^2[\phi] = \prod_{\vec{k}_F, \tau,\vec{q},\vec{q}
\cdot \vec{v}_{\vec{k}_F}>0}
\frac{d\phi_{\vec{q}}(\vec{k}_F,\tau)d\phi^{*}_{\vec{q}}(\vec{k}_F,\tau)}
{\pi N_{\Lambda}(\vec{k}_F) V |\vec{q} \cdot \vec{v}_{\vec{k}_F}|}$ and
\begin{equation}
i S[\phi] = \sum_{\vec{k}_F,\vec{q},\vec{q} \cdot \vec{v}_{\vec{k}_F}>0}
\frac{1}{N_{\Lambda}(\vec{k}_F) V
|\vec{q} \cdot \vec{v}_{\vec{k}_F}|}
\left[\frac{1}{2} \left(\tilde{\varphi}_{\vec{q}}(\vec{k}_F)
\phi^{*}_{\vec{q}}(\vec{k}_F,0)+
\varphi^{*}_{\vec{q}}(\vec{k}_F)
\phi_{\vec{q}}(\vec{k}_F,t)\right) + \int_{0}^{t} d\tau
L(\phi(\tau))\right]
\end{equation}
is the classical action for the motion of the bosonic fields and
\begin{equation}
L(\phi) = \sum_{\vec{k}_F} \frac{1}{2}\left(\phi_{\vec{q}}(\vec{k}_F) \frac{d
\phi^{*}_{\vec{q}}(\vec{k}_F)}{d\tau} -
\phi^{*}_{\vec{q}}(\vec{k}_F) \frac{d \phi_{\vec{q}}(\vec{k}_F)}
{d\tau}\right) - i \tilde{H}[\phi(\tau)]
\end{equation}
is the lagrangian density. The rescaled Hamiltonian $\tilde{H}$
is defined as
\begin{equation}
\tilde{H}[\phi(\tau)]= \sum_{\vec{k}_F}
N_{\Lambda}(\vec{k}_F)V |\vec{q} \cdot \vec{v}_{\vec{k}_F}|
\, \, \frac{\langle \vec{k}_F,[\phi]|H| \vec{k}_F,[\phi]\rangle}
{\langle \vec{k}_F,[\phi]|\vec{k}_F,[\phi]\rangle}.
\end{equation}
In the path integral the boundary conditions are defined by the
$S$-matrix (3.6),
\begin{eqnarray}
\phi_{\vec{q}}(\vec{k}_F,0) = \tilde{\varphi}_{\vec{q}}(\vec{k}_F)
\nonumber
\\
\phi^{*}_{\vec{q}}(\vec{k}_F,t)=\varphi^{*}_{\vec{q}}(\vec{k}_F).
\end{eqnarray}

As in any field theory we can also write a generating functional,
$Z$, which is the trace of the $S$-matrix. In terms of path integrals
it is written as,
\begin{equation}
Z = \int D^2[\phi]
e^{i \int dt \, \sum_{\vec{q},\vec{k}_F}
L(\phi^{*}_{\vec{q}}(\vec{k}_F,t)\phi_{\vec{q}}(\vec{k}_F,t))}
\end{equation}
where the Lagrangian is same as in (3.9) and (3.10). Now we have
periodic boundary conditions due to the trace.
As usual the Euler-Lagrange equations for the classical action
will generate the semiclassical dynamics for the problem. We have
shown that in case of a Fermi liquid the semiclassical dynamics
is represented by the Landau equation of sound waves. Therefore,
the bosonic fields represent these waves which are distortions
of the Fermi surface evolving in time. Moreover, the same functional
integral in imaginary time reproduces the thermodynamics of
the Fermi liquids \cite{us}.

As we will see in the next section these results, besides providing
a new physical insight into the physics of fermionic systems
(and an interpretation for the existence of bosons in a fermionic
theory), are also powerful tools which will allow us to calculate
correlation functions of interest.
To this end, however, we need a dictionary to translate
from the language of fermions to bosons. The aim of the next
section is to provide it.

\section{The fermion operator}

The connection between bosons and fermions is inspired from the bosonization
methods in one-dimensional systems. Luther \cite{luther} was the first
to explore the analogy between one
dimension and higher dimensions. Luther's idea was to define
at each point of the Fermi surface a one dimensional system using the
radial directions. However, Luther worked only with non-interacting fermions.
Here, we present an argument paralleling Luther's work.

The fermion operator $\psi(\vec{r})$ is written in momentum space as,
\begin{equation}
\psi(\vec{r}) = \frac{1}{\sqrt{V}} \sum_{\vec{k}} e^{i \vec{k} \cdot \vec{r}}
c_{\vec{k}}.
\end{equation}

Inspired by the results in one dimension systems we follow Luther's
construction of the fermionic operator $\Psi(\vec{r},\vec{k}_F)$ as follows,
\begin{eqnarray}
\psi(\vec{r}) = \sum_{\vec{k}_F} \Psi(\vec{r},\vec{k}_F),
\nonumber
\end{eqnarray}
where,
\begin{equation}
\Psi(\vec{r},\vec{k}_F) = f(\vec{k}_F) e^{J(\vec{r},\vec{k}_F)},
\end{equation}
for some operators $f(\vec{k}_F)$ and $J(\vec{r},\vec{k}_F)$.
The correct commutation relations for the $\Psi(\vec{r},\vec{k}_F)$
can be obtained by imposing the commutation relation
$[n_{\vec{q}}(\vec{k}),f(\vec{k}_F)]=0$ and
$[n_{\vec{q}}(\vec{k}),J(\vec{r},\vec{k}_F)]= c-number$.
This choice seems to be the simplest possible.

Since the operators $n_{\vec{q}}(\vec{k})$ generate the Hilbert space of
interest, the form of the fermion operator $\Psi(\vec{r},\vec{k}_F)$
will be
defined by the commutation relation between the former and the latter.
It is easy to show, using the fermionic algebra for the operators
$c_{\vec{k}}$, that,
\begin{equation}
[\psi(\vec{r}),\sum_{\vec{k}} n_{\vec{q}}(\vec{k})] = e^{-i \vec{q} \cdot
\vec{r}}
\psi(\vec{r}).
\end{equation}
{}From (4.2) we have,
\begin{equation}
[\Psi(\vec{r},\vec{k}_F),n_{\vec{q}}(\vec{k'}_F)] =
e^{-i \vec{q} \cdot \vec{r}}
\Psi(\vec{r},\vec{k'}_F) \, \delta_{\vec{k}_F,\vec{k'}_F}.
\end{equation}

Using the choices for the commutation relations between the operators
$f(\vec{k}_F)$ and $J(\vec{r},\vec{k}_F)$ with $n_{\vec{q}}(\vec{k})$,
we easily get,
\begin{equation}
[\Psi(\vec{r},\vec{k}_F),n_{\vec{q}}(\vec{k}_F)] = [n_{\vec{q}}(\vec{k}_F),
J(\vec{r},\vec{k}_F)] \Psi(\vec{r},\vec{k}_F).
\end{equation}
Therefore, the fermion operator will be well defined if we insure that,
\begin{equation}
[n_{\vec{q}}(\vec{k}_F),J(\vec{r},\vec{k}_F)] = e^{-i \vec{q} \cdot \vec{r}}
\end{equation}
Due to the commutation relations (2.2) it is easy to see that
the correct choice is,
\begin{equation}
J(\vec{r},\vec{k}_F)= - \sum_{\vec{q}} \frac{e^{-i \vec{q} \cdot \vec{r}}}
{N_{\Lambda}(\vec{k}_F) V\vec{q} \cdot \vec{v}_{\vec{k}_F}}
n_{-\vec{q}}(\vec{k}_F).
\end{equation}

Notice the similarity between the form (3.1) and (4.7). This is not
a mere coincidence.
The fermion operator is represented as a coherent state of bosons.
Since, as we will see, the bosons diagonalize the problem of an interacting
electronic system, the fermion operator is a non-perturbative object in
the language of the bosons.
Moreover, since the operator $U$ is a functional of the fields we observe
that the fermion operator $\Psi(\vec{r},\vec{k}_F)$
is given by $U$ for the following choice of the fields
\begin{equation}
\phi_{\vec{q}}(\vec{k'}_F)=e^{-i \vec{q} \cdot \vec{r}}
\delta_{\vec{k}_F,\vec{k'}_F} ,
\end{equation}
which means that while the fermions interact among themselves the
bosons are free excitations in the Fermi surface.

{}From this observation we conclude that the fermion propagator,
\begin{equation}
K(\vec{r}-\vec{r'},t-t') = \sum_{\vec{k}_F,\vec{k'}_F}
\langle FS|\Psi^{\dag}(\vec{r},\vec{k}_F,t) \Psi(\vec{r'},\vec{k}_F,t')
|FS \rangle
\end{equation}
can be written as,
\begin{equation}
K(\vec{r}-\vec{r'},t-t') = \sum_{\vec{k}_F,\vec{k'}_F}
f^{*}(\vec{k}_F) f(\vec{k'}_F)
\langle \vec{k}_F,\vec{r},t| \vec{k'}_F,\vec{r'},0\rangle,
\end{equation}
where $|\vec{k}_F,\vec{r},t\rangle$ is the coherent state (3.4) with the
prescription (4.8).
The functions $f(\vec{k})$ are written in terms of the total number of
fermions in the system \cite{luther,halpern}. Their role is to insure that the
fermion operators anticommute with each other.

The one-particle Green's function can be obtained directly
from the propagator in the case when the actual ground state
of the system is the filled Fermi sea:
\begin{equation}
G(\vec{r}-\vec{r'},t-t')= K(\vec{r}-\vec{r'},t-t') \Theta(t-t')
-K(\vec{r}-\vec{r'},t-t')^{*} \Theta(t'-t).
\end{equation}
As we will show in the next section this is the case of the
non-interacting electronic system and Fermi liquids.

\section{The Green's function for the free system}

The hamiltonian for a non-interacting electronic system is
just the kinetic term,
\begin{equation}
H_0 = \sum_{\vec{k}} \epsilon_{\vec{k}} c^{\dag}_{\vec{k}} c_{\vec{k}}.
\end{equation}

Since the operators (2.1) generate the Hilbert space of interest,
the form of the hamiltonian (5.1) in this space will depend on
the commutation relations between the hamiltonian and these operators.
It is easy to show that \cite{us},
\begin{equation}
\left[H_0,b_{\vec{q}}(\vec{k}_F)\right] = -|\vec{q} \cdot\vec{v}_{\vec{k}_F}|
b_{\vec{q}}(\vec{k}_F),
\end{equation}
therefore the non-interacting hamiltonian can be written as,
\begin{equation}
H_0 = \sum_{\vec{k}_F} \sum_{\vec{q},\vec{q} \cdot \vec{v}_{\vec{k}_F}>0}
|\vec{q} \cdot \vec{v}_{\vec{k}_F}|
b^{\dag}_{\vec{q}}(\vec{k}_F) b_{\vec{q}}(\vec{k}_F).
\end{equation}
This result is only valid in the restricted Hilbert space of
states which lie close to the Fermi surface.

Therefore, the free system, which is composed by only the continuum
of particle-hole pairs, is described by a set of independent harmonic
modes at each point of the Fermi surface. These modes oscillate
with arbitrary phase difference between them. It is an incoherent
oscillation of the Fermi surface which vanishes on average. It
means that, on average, the shape of the Fermi surface is kept
constant, as expected.

Given the hamiltonian (5.3) we can use the methods of the last section
in order to calculate the non-interacting Green's function.
The rescaled hamiltonian defined in (3.10) is written as,
\begin{equation}
\tilde{H}[\phi(\tau)]= |\vec{q} \cdot \vec{v}_{\vec{k}_F}|
\phi^{*}_{\vec{q}}(\vec{k}_F) \phi_{\vec{q}}(\vec{k}_F).
\end{equation}

The functional integral we obtain is the same as for the
harmonic oscillator and it is easily done \cite{klauder},
\begin{eqnarray}
\langle \vec{k}_F,[\varphi],t| \vec{k'}_F,[\tilde{\varphi}],0\rangle =
\exp \left\{  - \sum_{\vec{q},\vec{q} \cdot \vec{v}_{\vec{k}_F}>0} \,
\left(\frac{1}{2 N_{\Lambda}(\vec{k}_F) V}\right) \,
\left[\frac{1}{|\vec{q} \cdot
\vec{v}_{\vec{k}_F}|}|\varphi_{\vec{q}}(\vec{k}_F)|^2 +
\frac{1}{|\vec{q} \cdot \vec{v}_{\vec{k'}_F}|}
|\tilde{\varphi}_{\vec{q}}(\vec{k'}_F)|^2 \right. \right.
\nonumber
\\
\left. \left. - \frac{2}{|\vec{q} \cdot \vec{v}_{\vec{k}_F}|}
\delta_{\vec{k}_F,\vec{k'}_F} e^{-i|\vec{q} \cdot \vec{v}_{\vec{k}_F}|t}
\tilde{\varphi}_{\vec{q}}(\vec{k}_F)
\varphi^{*}_{\vec{q}}(\vec{k}_F) \right] \right\}.
\end{eqnarray}

We have seen that the propagator is obtained in this language by
using the following choice (4.8),
\begin{equation}
\varphi_{\vec{q}}(\vec{k}_F)=
e^{-i \vec{q} \cdot \vec{r} - \frac{\alpha |\vec{q} \cdot \vec{v}_{\vec{k}_F}|}
{2|\vec{v}_{\vec{k}_F}|}}
\end{equation}
where $\alpha$ ($\sim \Lambda^{-1}$) is a cut-off in the normal direction
to the Fermi surface
which we introduce in order to regularize the radial integrals (exactly
as in one dimension).
Making the substitution above we find,
\begin{eqnarray}
\langle \vec{k}_F,\vec{r},t| \vec{k'}_F,0,0\rangle_0 =
\exp \left\{ -\sum_{\vec{q},\vec{q} \cdot \vec{v}_{\vec{k}_F}>0} \,
\left(\frac{1}{2 N_{\Lambda}(\vec{k}_F) V}\right) \,
\left[ \frac{1}{|\vec{q} \cdot \vec{v}_{\vec{k}_F}|} e^{-\frac{\alpha |\vec{q}.
\vec{v}_{\vec{k}_F}|}{|\vec{v}_{\vec{k}_F}|}}+
\frac{1}{|\vec{q}.\vec{v}_{\vec{k'}_F}|} e^{-\frac{\alpha |\vec{q}.
\vec{v}_{\vec{k'}_F}|}{|\vec{v}_{\vec{k'}_F}|}}  \right. \right.
\nonumber
\\
\left. \left. - \frac{2 \delta_{\vec{k}_F,\vec{k'}_F}}{|\vec{q} \cdot
\vec{v}_{\vec{k}_F}|}
e^{-\frac{\alpha |\vec{q} \cdot \vec{v}_{\vec{k}_F}|}{|\vec{v}_{\vec{k}_F}|}}
e^{i(\vec{q} \cdot \vec{r} - |\vec{q} \cdot \vec{v}_{\vec{k}_F}|t)} \right]
\right\}.
\end{eqnarray}

Notice that for different points of the Fermi surface the integral
diverges logarithmically due to the factor $1/|\vec{q}.\vec{v}_{\vec{k'}_F}|$
which diverges in the the limit of $q \to 0$. Thus, we can write,
\begin{equation}
\langle \vec{k}_F,\vec{r},t| \vec{k'}_F,0,0\rangle_0 = 0
\hspace{1cm} \vec{k}_F \neq \vec{k'}_F.
\end{equation}

If $\vec{k}_F=\vec{k'}_F$ we find the following integral,
\begin{equation}
\langle \vec{k}_F,\vec{r},t| \vec{k'}_F,0,0\rangle_0 =
\delta_{\vec{k}_F,\vec{k'}_F}
\exp\left(-\sum_{\vec{q},\vec{q} \cdot \vec{v}_{\vec{k}_F}>0}
 \frac{1}{N_{\Lambda}(\vec{k}_F) V}
\left(1-e^{i(\vec{q} \cdot \vec{r} -|\vec{q} \cdot
\vec{v}_{\vec{k}_F}|t)}\right)
\frac{e^{-\frac{\alpha |\vec{q} \cdot \vec{v}_{\vec{k}_F}|}
{|\vec{v}_{\vec{k}_F}|}}}{|\vec{q} \cdot \vec{v}_{\vec{k}_F}|}\right).
\end{equation}

In order to evaluate this integral we observe that the
component of $\vec{q}$ normal to the Fermi surface dominates the
behavior of the integral. It is natural, therefore,
to split $\vec{q} =\vec{q_N}+\vec{q_T}$ where $\vec{q_N}=
(\vec{q}.\vec{n}_{\vec{k}_F}) \vec{n}_{\vec{k}_F}$ (with $\vec{n}_{\vec{k}_F}=
\frac{\vec{v}_{\vec{k}_F}}{|\vec{v}_{\vec{k}_F}|}$) which is the normal
component to the Fermi surface and the tangential component $\vec{q_T}$
which is defined as $\vec{q_T}.\vec{n}_{\vec{k}_F} =0$. This choice
can be viewed as a Fresnel construction of differential geometry.
We have built a local reference frame on the Fermi surface which
follows its local geometry.

Observe that the integrals over the tangent component can be
easily evaluated. They have the form,
\begin{equation}
\int_{-\infty}^{\infty} dq_T q^{d-2}_T e^{i q_T x_T} = \frac{1}{i^{d-2}}
\frac{\partial^{d-2}}{\partial x^{d-2}_T} \left(\delta(x_T))\right)
\end{equation}
which gives a negligible contribution except for $x_T = 0$. We conclude,
therefore, that the part tangent to the Fermi surface does not contribute
to the long distance behavior of the Green's function and therefore
can be neglected in this limit. However, it does contribute to the
counting
of states at the Fermi surface. Indeed, from (2.10), we see that the local
density of states can be written as,
\begin{equation}
N_{\Lambda}(\vec{k}_F) = \frac{1}{V} \int
\frac{d\vec{S}}{|\vec{v}_{\vec{k}_F}|}
|\Phi_{\Lambda}(\vec{k}_F)|^2
\end{equation}
where  $d\vec{S}$ is the area element on the Fermi surface. However, the area
on the surface is written as an integral over the tangential component
of
$\vec{q}$ alone. Therefore, the tangential part of the integral
contributes
to the density of states and not to the long distance behavior of
the correlation functions. Thus we make the following substitution,
\begin{equation}
\sum_{\vec{q},\vec{q} \cdot \vec{v}_{\vec{k}_F}>0} \to
N_{\Lambda}(\vec{k}_F) V |\vec{v}_{\vec{k}_F}| \int_{0}^{\infty} dq_N.
\end{equation}
The same type of substitution gives the correct thermodynamic properties
for these systems, namely, a linear specific heat proportional to the
density of states \cite{us}.

Substituting (5.12) in (5.9) we finally get,
\begin{equation}
\langle \vec{k}_F,\vec{r},t| \vec{k'}_F,0,0\rangle_0 =
\delta_{\vec{k}_F,\vec{k'}_F}
\exp\left(-\int_{0}^{\infty} dq_N \left(1-
e^{i q_N (\vec{n}_{\vec{k}_F}.\vec{r}-|\vec{v}_{\vec{k}_F}|t)}\right)
\frac{e^{-\alpha q_N}}{q_N}
\right).
\end{equation}
Now we use the integral,
\begin{equation}
\int_{0}^{\infty} dx \left(1-e^{i x a}\right) \frac{e^{-\alpha x}}{x}
= \ln\left(1- i \frac{a}{\alpha}\right),
\end{equation}
and we finally conclude,
\begin{equation}
\langle \vec{k}_F,\vec{r},t| \vec{k'}_F,0,0\rangle_0 = \frac{i \alpha
\delta_{\vec{k}_F,\vec{k'}_F}}{\vec{n}_{\vec{k}_F}.\vec{r}-
|\vec{v}_{\vec{k}_F}|t + i\alpha}
\end{equation}
which is the correct asymptotic form for the non-interacting system in
any number of dimensions \cite{luther}. It represents a fermion moving
with velocity $\vec{v}_{\vec{k}_F}$, as expected.

\section{The Hamiltonian}

We now discuss the possible forms of the hamiltonian
which can appear in the problem. We assume that the hamiltonian
is composed of two terms, the kinetic energy of the fermions, eq.~(5.1),
and a two-particle interaction,
\begin{equation}
U =\frac{1}{2 V}
\sum_{\vec{p},\vec{p'},\vec{q}} \, U_{\vec{p},\vec{p'}}(q)
\, \, \, c^{\dag}_{\vec{p}+\frac{\vec{q}}{2}} c_{\vec{p}-\frac{\vec{q}}{2}}
c^{\dag}_{\vec{p'}-\frac{\vec{q}}{2}} c_{\vec{p'}+\frac{\vec{q}}{2}}.
\end{equation}
The scattering processes described by the operators in (6.1) can be visualized
by the pairing of the annihilation and creation operators which are depicted
in Fig.1-a and Fig.1-b. In Fig.1-a, we have small momentum transfer
$\vec{q}$ (forward scattering) between the pairs. In Fig.1-b the
pairs exchange momentum $\vec{p}-\vec{p'}$ can be of order of
$2 k_F$ (backward scattering).

The form of the kinetic energy in terms of the bosonic operators
was discussed in the previous section and it is given by Eq.(5.3).
In terms of the operators which generate the Hilbert space of interest,
it is easy to see that the interaction can be written as,
\begin{equation}
U=\frac{1}{2 V} \sum_{\vec{p},\vec{p'},\vec{q}} \, U_{\vec{p},\vec{p'}}(q) \,
\, n_{-\vec{q}}(\vec{p}) n_{\vec{q}}(\vec{p'}).
\end{equation}

We can now parametrize the form of the hamiltonian in terms of
processes involving particle-hole pairs close to the Fermi surface.
 From (2.5) we get four kind
of terms (these are the {\it only} types of terms (scattering processes)
which are present in the restricted Hilbert space),
\begin{eqnarray}
a^{\dag}_{\vec{q}}(\vec{k}_F) a_{\vec{q}}(\vec{k'}_F)
\\
\vspace{1cm} \, \, for \, \, \, \vec{q} \cdot \vec{v}_{\vec{k}_F}>0 \, , \,
\vec{q} \cdot \vec{v}_{\vec{k'}_F}>0;
\nonumber
\end{eqnarray}
\begin{eqnarray}
a_{\vec{q}}(\vec{k}_F) a^{\dag}_{\vec{q}}(\vec{k'}_F)
\\
\vspace{1cm} \, \, for \, \, \, \vec{q} \cdot \vec{v}_{\vec{k}_F}<0 \, , \,
\vec{q} \cdot \vec{v}_{\vec{k'}_F}<0;
\nonumber
\end{eqnarray}
\begin{eqnarray}
a_{\vec{q}}(\vec{k}_F) a_{\vec{q}}(\vec{k'}_F)
\\
\vspace{1cm} \, \, for \, \, \, \vec{q} \cdot \vec{v}_{\vec{k}_F}<0 \, , \,
\vec{q} \cdot \vec{v}_{\vec{k'}_F}>0;
\nonumber
\end{eqnarray}
\begin{eqnarray}
a^{\dag}_{\vec{q}}(\vec{k}_F) a^{\dag}_{\vec{q}}(\vec{k'}_F)
\\
\vspace{1cm} \, \, for \, \, \, \vec{q} \cdot \vec{v}_{\vec{k}_F}>0 \, , \,
\vec{q} \cdot \vec{v}_{\vec{k'}_F}<0.
\nonumber
\end{eqnarray}
For a fixed direction of $\vec{q}$ the processes (6.3), (6.4) and (6.5),
(6.6) are shown in Fig.2 and Fig.3, respectively.

Observe that the momentum $\vec{q}$ breaks the rotation symmetry O(3) of
the Fermi surface (for the isotropic system, of course) and
introduces an axis in the problem. We can now divide the Fermi
sea into  ``south"and ``north" hemispheres. While the
interactions (6.3) and (6.4) occur within one of the hemispheres
the interactions (6.5) and (6.6) link different hemispheres of
the Fermi sea.

In Fermi liquid theory only the interactions (6.3) and (6.4)
are present \cite{pines}.
The reason for that, in the absence of nesting and singular
interactions, is very easy to
understand. The scattering of a particle-hole pair initially
located at vector $\vec{k}_F$ to a new position $\vec{k'}_F$
requires an amount of energy $|\vec{q} \cdot
\left(\vec{v}_{\vec{k}_F}-\vec{v}_{\vec{k'}_F}\right)|$. Since in
perturbation theory the denominators in the perturbative expansion
are written as $\omega-\vec{q} \cdot
\left(\vec{v}_{\vec{k}_F}-\vec{v}_{\vec{k'}_F}\right)$,
at low energy, the most important contribution, for
fixed momentum transfer comes from  regions of momentum space
which $\vec{v}_{\vec{k}_F}
\sim \vec{v}_{\vec{k'}_F}$. This is only true, of course, if
the matrix elements between the states considered are not singular.
Indeed, it is worth noticing that
the corrections to the specific heat due to the interaction between
particle-hole pairs is calculated in this limit \cite{Baym,us,carneiro}
and they agree very well with the experimental results. The presence
of this process will give rise to terms of the
form $a^{\dag} \, a$ in the hamiltonian. That
is, the same form as the non-interacting system (except for being
non-diagonal in the index $\vec{k}$) \cite{us}. Therefore, these interactions
never mix the creation and annihilation operators for the bosons.
As a consequence, the only effect that these interactions can produce
is a different phase of oscillation for the fields (see section
IX below), or, in other words, to renormalize the Fermi velocity.

However, the terms (6.5) and (6.6) which link the ``south" and ``north"
hemispheres are more interesting. They will appear in the Hamiltonian as a
combination of $a^{\dag} \, a^{\dag}$ and $a\,a$. The consequence
of these terms is dramatic. Besides changing the phase of oscillation,
they also mix creation and annihilation operators. It means that they
introduce a renormalization of the amplitude of the fields by a
kind of Bogoliubov transformation. If this process survives in the
thermodynamic limit it can be shown to give rise
to anomalous dimensions in the correlations functions.
We notice here that this is the same kind of interaction
which produces anomalous dimensions in one-dimension.

Indeed, suppose we have the interacting
part of the hamiltonian written for an one dimensional lattice as,
\begin{equation}
H= U \sum_{n} \rho_n \, \rho_{n+1}
\end{equation}
where $\rho_n$ is the charge density at the site $n$. The interacting
term (6.7) gives rise to two different terms in the bosonized
hamiltonian \cite{bosonization},
\begin{equation}
U \left(\rho_1(q) \rho_1(-q) + \rho_2(q) \rho_2(-q)\right)
\end{equation}
\begin{equation}
2 U \rho_1(q) \rho_2(-q) ,
\end{equation}
where $\rho_1(q)$ and $\rho_2(q)$ are the Fourier component of
the ``right" and ``left" movers, respectively (they are analogous
of the ``south" and ``north" movers).

The interaction term (6.8) is associated with the processes described
by (6.3) and (6.4) which are responsible for forward scattering.
The process (6.9) is related to the terms (6.5) and (6.6).
The same effect will occur in higher dimensions,
for analogous reasons, if we have a nested Fermi surface.
Observe that in the case of nesting the vectors $\vec{v}_{\vec{k}_F}$
at one side of Fermi surface are the same and therefore processes
which translate the particle-hole pair within one side of the Fermi
surface cost no energy ($|\vec{q} \cdot \left(\vec{v}_{\vec{k}_F}-
\vec{v}_{\vec{k'}_F}\right)| = 0$). The only other available process
in the system transfer particle-hole pairs across the
Fermi sea. They are described by (6.5) or (6.6) (or (6.9) in
the one dimensional case which is a special case of nesting). This
process will appear in the case of a nested Fermi surface even
in the absence of singular matrix elements.
We note here that the vanishing of the quasiparticle residue
due to nesting was already obtained in numerical calculations
in the two-dimensional Hubbard model close to half filling
\cite{guinea}.

Notice also that the processes in different hemispheres of the
Fermi surface require an amount of energy ($ |\vec{q}
\cdot \left(\vec{v}_{\vec{k}_F}-\vec{v}_{\vec{k'}_F}\right)|$)
which, for fixed momentum transfer, does not vanish because
$\vec{v}_{\vec{k}_F}\sim -\vec{v}_{\vec{k'}_F}$. Therefore, these
processes will be marginally irrelevant at low
energy and will not contribute for the physics of the states
close of the Fermi surface. Indeed, by a perturbative renormalization
group approach to Fermi liquids, Shankar \cite{shankar} has shown
explicitly the marginality of these processes. In other words, in normal
conditions
these processes will scale away in the thermodynamic limit
in the bosonic form of the hamiltonian. We will show this
result explicitly in the next section with the exact diagonalization
of the bosonic hamiltonian.
For these processes to become part of the bosonic hamiltonian
special physical interactions are needed.
For instance, even in perturbation
theory, these processes will become important if the matrix
elements between the states is singular at low energies and
long wavelengths. It is clear that in this case the perturbative
expansion fails. More importantly, it is possible to show that coupling
a Fermi liquid to dynamical gauge fields {\it generically} leads to
interactions of this type and, consequently, to a breakdown of Fermi
liquid theory. We will
study this problem using our methods in a forthcoming paper \cite{next}.

\section{Diagonalization of the bosonic hamiltonian}

As it was explained in the previous section the hamiltonian for
interacting fermions close to the Fermi surface can be rewritten
in bosonic form as,
\begin{eqnarray}
H = H_0 + \sum_{\vec{k}_F, \vec{k}_{F'}, \vec{q}}
\sqrt{\vec{q} \cdot \vec{v}_{\vec{k}_F} \vec{q} \cdot \vec{v}_{\vec{k}_{F'}}
N_{\Lambda}(\vec{k}_F) N_{\Lambda}(\vec{k}_{F'})} U(q) \left(
b^{\dag}_{\vec{q}}(\vec{k}_F) b_{\vec{q}}(\vec{k}_{F'}) +
b^{\dag}_{\vec{q}}(-\vec{k}_F) b_{\vec{q}}(-\vec{k}_{F'})+ \right.
\nonumber
\\
\left. b_{\vec{q}}(\vec{k}_F) b_{\vec{q}}(-\vec{k}_{F'}) +
b^{\dag}_{\vec{q}}(-\vec{k}_{F'}) b^{\dag}_{\vec{q}}(\vec{k}_{F})
\right)
\end{eqnarray}
where the sum is restricted to
vectors such that $\vec{q} \cdot \vec{v}_{\vec{k}_F}>0$ and
$\vec{q} \cdot \vec{v}_{\vec{k}_{F'}}>0$. We also have used the
definition (2.12) and (2.5) and changed the
sums to be restricted to just one hemisphere of the Fermi surface.
The free term, $H_0$ is already defined in (5.3) and we assume that
the Fermi surface is round and the interaction is isotropic (these
assumptions are made in order to simplify the calculations, however
none of them are really important in what follows).

In order to make notation simpler we discretize the hamiltonian in
such a way that the Fermi sphere is divided into $N$ patches \cite{houghton}
with area $S_{d-1} \Lambda^{d-1}$ such that the total area of the
Fermi sphere is written as $S_d k_F^{d-1}=S_{d-1} \Lambda^{d-1} N$.
{}From this definition the local density of states at the Fermi points
is simply $N_{\Lambda}(\vec{k}_F) = N(0)/N$ where the overall density of
states is $N(0) = \frac{S_d k_F^{d-1}}{(2 \pi)^d v_F}$.
In this case the dimensionless coupling constant of the theory is given by,
\begin{equation}
g(q) = N_{\Lambda}(\vec{k}_F) U(q) = \frac{N(0) U(q)}{N} = \frac{S_{d-1}
\Lambda^{d-1} U(q)}{(2 \pi)^d N}
\end{equation}
which vanishes in the thermodynamic limit for dimensions greater than one.
This is the route for the stability of Fermi liquid theory as already
discussed by Haldane \cite{haldane,houghton,us,shankar}.

{}From now on we consider the Fermi liquid problem in two dimensions, partly
because of its relevance of the cuprates and because of its general interest
\cite{anderson,ian}. To each point $j$ in the Fermi circle there is an angle
$\theta_j = \frac{2 \pi}{N} j$. Since in the sum in (7.1) is restricted
to $\vec{q} \cdot  \vec{v}_{\vec{k}_F}>0$ the values of $j$ the allowed
values of $j$ are $-\frac{N}{4}+1 \leq j \leq \frac{N}{4}-1$. In the
direction of $\vec{q}$ we have $j=0$. In this notation we can rewrite,
\begin{equation}
\vec{q} \cdot  \vec{v}_{\vec{k}_j} = q v_F \cos(\theta_j).
\end{equation}
Moreover, we define the following notation for the bosonic operators,
\begin{eqnarray}
b_{\vec{q}}(\vec{k}_j) = b_j
\nonumber
\\
b_{\vec{q}}(-\vec{k}_j) = a_j.
\end{eqnarray}

Using the previous definitions it is easy to see that the hamiltonian
(7.1) can be rewritten in the following form,
\begin{equation}
H = \sum_{\vec{q}} q v_F {\cal H}(q)
\end{equation}
where
\begin{equation}
{\cal H} = \sum_{i} s_i \left(b^{\dag}_i b_i + a^{\dag}_i a_i\right)
+ g \sum_{i,j} \sqrt{s_i s_j} \left(b^{\dag}_i b_j
+ a^{\dag}_i a_j + a_i b_j + b^{\dag}_i a^{\dag}_j \right),
\end{equation}
where $s_i = \cos(\theta_i)$. In the last expression we have dropped
the index $q$ since the original hamiltonian is already diagonal in this
index. Observe that the hamiltonian (7.6) describe a set of coupled
harmonic oscillators. In order to diagonalize this hamiltonian we
define a generalized Bogoliubov transformation which mixes different
points at the Fermi surface. We introduce two real orthogonal matrices
${\cal M}_{il}$ and ${\cal N}_{il}$ and two new bosonic operators
$\beta_l$ and $\alpha_l$ such that,
\begin{eqnarray}
b_i = \sum_{l} \left({\cal M}_{il} \beta_l + {\cal N}_{il} \alpha_l^{\dag}
\right)
\nonumber
\\
a_i = \sum_{l} \left({\cal M}_{il} \alpha_l + {\cal N}_{il} \beta_l^{\dag}
\right).
\end{eqnarray}

The commutation relation between the operators enforces that,
\begin{eqnarray}
\left[b_i,b^{\dag}_j\right]=\left[a_i,a^{\dag}_j\right]=\delta_{i,j} =
\sum_{l} \left({\cal M}_{il} {\cal M}_{jl}-{\cal N}_{il} {\cal N}_{jl}\right)
\nonumber
\\
\left[b_i,a_j\right] = 0 = \sum_{l} \left({\cal M}_{il} {\cal N}_{jl}
-{\cal N}_{il} {\cal M}_{jl}\right),
\end{eqnarray}
where the new bosonic operators obey the usual algebra,
$\left[\beta_l,\beta_k^{\dag}\right] = \left[\alpha_l,\alpha_k^{\dag}\right]
= \delta_{l,k}$ and $\left[\beta_l,\beta_k\right] = \left[\alpha_l,
\alpha_k\right]=0$.

It is assumed that these new operators diagonalize the problem
completely, that is, the hamiltonian is written now as,
\begin{equation}
{\cal H} = \sum_l S_l
\left(\beta_l^{\dag} \beta_l + \alpha_l^{\dag} \alpha_l\right),
\end{equation}
where $S_l$ is the eigenfrequency.

To find the equation which defines the above matrices we look at
the commutation relation between the hamiltonian and the operators.
Using the hamiltonian (7.6) and the definition (7.7) with the
hamiltonian (7.9) we find,
\begin{equation}
\left[b_i,{\cal H}\right]= s_i b_i + g \sum_j \sqrt{s_i
s_j} \left(b_j + a^{\dag}_j\right) = \sum_{l} S_l \left({\cal M}_{il}
\beta_l - {\cal N}_{il} \alpha_l^{\dag}\right).
\end{equation}
Substituting (7.7) and taking the commutation relations with the
operators $\alpha_l$ and $\beta_l$ we find the following equations,
\begin{eqnarray}
\left(S_l-s_i\right){\cal M}_{il} = g \sum_j \sqrt{s_i
s_j} \left({\cal M}_{jl}+{\cal N}_{jl}\right)
\nonumber
\\
\left(S_l+s_i\right){\cal N}_{il} = - g \sum_j \sqrt{s_i
s_j} \left({\cal M}_{jl}+{\cal N}_{jl}\right).
\end{eqnarray}
These equations define the Bogoliubov transformation.

Notice that in principle we have two different kinds of solutions:
the first one is related to the particle-hole continuum and it
is defined for $S_l \leq 1$ and the second is related to the
collective modes and is obtained for $S_l > 1$. All the
solutions of the particle-hole continuum can be written in the
form $S_l = \cos(\theta_l + \delta_l)$ where $ \delta_l$
is a correction of order $\frac{1}{N}$ and they correspond to
an angular shift of the positions of the points at the Fermi surface.
In the thermodynamic limit (which is the limit of interest) they
fill densely the Fermi surface and renormalize to the bare frequencies
\cite{pines}. The collective modes cannot be written as a cosine of
some angle. They detach from the particle-hole continuum and represent
high energy excitations which are separated from it by an energy gap
which depends on the interaction. However, in principle, they are
also present in the solution of the bosonic hamiltonian. In the case
of the Landau theory there is only one collective mode (the zero sound
for neutral systems or the plasmon for charged systems) which is
born at the direction of the momentum transfer $\vec{q}$, or in
our notation, at $j=0$. In some sense the Fermi liquid problem resembles
an impurity problem in which the states of the particle-hole
continuum get only a phase shift and a bound state (the collective mode)
appears in the spectrum. We therefore divide our solution in two
cases as follows.

\subsection{Particle-hole continuum $S_l=s_l$}

Observe that in this case the equation (7.11) can only be inverted
if we disappear with the singularity which appears for $i=l$.
this is done be writing (7.11) as,
\begin{eqnarray}
{\cal M}_{il} = Z_l g \sqrt{s_l} C_l \delta_{i,l} +
\frac{g \sqrt{s_i} C_l}{s_l-s_i}
\nonumber
\\
{\cal N}_{il} = - \frac{g \sqrt{s_i} C_l}{s_l+s_i},
\end{eqnarray}
where,
\begin{equation}
C_l = \sum_j \sqrt{s_j} \left({\cal M}_{jl}+{\cal N}_{jl}\right).
\end{equation}
The first term in the expression for ${\cal M}_{il}$ represents the
solution for $i=l$ where we introduced an unknown factor $Z_l$ which
must be evaluated. The second term must be understood as the principal
value of the function, that is, it vanishes for $i=l$. We must comment
here that we also need to avoid the term $i=-l$ since $s_l=
s_{-l}$. However, this only introduces simple modifications in
the algebra and does not affect the content of the results. It is
easy to show that with the introduction of these terms the final
result will be given in terms of symmetric and antisymmetric
combinations of the matrices that we obtain. We will come back to
this point latter in the paper.

By substituting the expression (7.12) in (7.13) we end up with an equation
for $Z_l$ which can be written as,
\begin{equation}
Z_l = \frac{1}{g s_l} \left(1-g \sum_j \frac{2 s_j^2}{
s_l^2-s_j^2}\right).
\end{equation}

\subsection{Collective mode $S_l = S_0$}

In the case of the collective mode there is no divergence in the
equation (7.11) and the solution is just,
\begin{eqnarray}
{\cal M}_{i0}= \frac{g \sqrt{s_i} C_0}{S_0-s_i}
\nonumber
\\
{\cal N}_{i0} = - \frac{g \sqrt{s_i} C_0}{S_0+s_i},
\end{eqnarray}
where $C_0$ is defined as in (7.13). By substituting these equations
there we find,
\begin{equation}
1 = g \sum_j \frac{2 s_j^2}{S_0^2-s_j^2},
\end{equation}
which defines $S_0$. The solution of this equation can be seen graphically
in fig.4. For a finite number of points the solutions for the particle-hole
continuum have a finite angular shift while the collective mode detaches
from the continuum. In the thermodynamic limit the particle-hole continuum
renormalizes to the bare frequencies while the collective mode gets a
finite renormalization in the frequency.

The sum in (7.16) can be rewritten in a well known form if we go back
to our original notation and take the thermodynamic limit ($N \to \infty$).
Using the definition of the beginning of the section we find,
\begin{eqnarray}
g \sum_j \frac{2 s_j^2}{s^2-s_j^2} = N(0) U(q)
\int_{-\pi/2}^{\pi/2} \frac{d\theta}{2 \pi} \frac{2 \cos^2(\theta)}
{s^2-\cos^2(\theta)} =
\nonumber
\\
 = N(0) U(q)  \int_{0}^{2 \pi}
\frac{d\theta}{2 \pi} \frac{\cos(\theta)}{s-\cos(\theta)} = U(q)
\Pi (s)
\end{eqnarray}
where,
\begin{equation}
\Pi\left(\frac{\omega}{q v_F}\right) = N(0) \int \frac{d\Omega}{S_d}
\frac{\vec{q} \cdot \vec{v}_{F}}{\omega-\vec{q} \cdot \vec{v}_{F}}
\end{equation}
is the RPA polarization function \cite{Baym}. Therefore, equation (7.16)
is nothing but the equation for the collective modes, as expected.
In two dimensions the form of the polarization function is given by
\cite{us},
\begin{equation}
\Pi(s) = N(0) \left(\frac{|s|}{\sqrt{s^2-1}} \Theta(|s|-1) +
\frac{s}{i \sqrt{1-s^2}}
\Theta(1-|s|) - 1 \right),
\end{equation}
where we have introduced a small imaginary part in the denominator in (7.18).
Notice that we can solve equation (7.16) for $S_0$ using (7.19) and (7.17).
Using the real part of (7.19) (the principal value) one finds
$S_0(q) = \frac{1+N(0)U(q)}{\sqrt{1+2 N(0)U(q)}}$ which is always
greater than $1$.

The normalization coefficients, $C_l$, in (7.12) are already undefined.
In order to calculate these coefficients we go back to the equations
(7.8). Using the results (7.12) the first of those equations gives,
\begin{eqnarray}
\delta_{i,j} = \delta_{i,j} Z_i^2 C_i^2 g^2 s_i +
g^2 \sqrt{s_i s_j} \left( s_i+s_j\right)
\left(\frac{Z_i C_i^2-Z_j C_j^2}{s_i^2-s_j^2} +
\right.
\nonumber
\\
\left. \sum_l \frac{ 2 C_l^2 s_l}{\left(s_l^2-s_j^2\right)
\left(s_l^2-s_i^2\right)}
+ \frac{ 2 C_0^2 S_0}{\left(S_0^2-s_j^2\right)
\left(S_0^2-s_i^2\right)}\right),
\end{eqnarray}
while the second gives,
\begin{equation}
\frac{Z_j C_j^2-Z_i C_i^2}{s_i^2-s_j^2} = \sum_l \frac{ 2 C_l^2
s_l}{\left(s_l^2-s_j^2\right)\left(s_l^2-s_i^2\right)}
+ \frac{ 2 C_0^2 S_0}{\left(S_0^2-s_j^2\right)
\left(S_0^2-s_i^2\right)}.
\end{equation}
By substitution of (7.21) in (7.20) we find finally that,
\begin{equation}
C_i = \frac{1}{Z_i g \sqrt{s_i}}.
\end{equation}

We can further simplify considerably our results in the thermodynamic limit.
The only sum which appears in the final form of the matrices ${\cal M}_{il}$
and ${\cal N}_{il}$ is the one related to the polarization function (7.19).
It is easy to show that,
\begin{equation}
\sum_j \frac{2 s_j^2}{s_l^2-s_j^2} = \frac{N}{2 \pi}
\left(s_l \int_{0}^{2 \pi} \frac{d\theta}{s_l-\cos \theta}
- 2 \pi\right) = -N,
\end{equation}
since the principal value of the integral in (7.23) vanishes for $|s_l|<1$
(see equation (7.19)).
If we substitute these results in (7.12) one gets,
\begin{eqnarray}
{\cal M}_{il} = \delta_{i,l} + \frac{1}{N} \; \frac{Ng}{1+Ng}
\; \; \frac{\sqrt{s_i s_l}}{s_l-s_i}
\nonumber
\\
{\cal N}_{il} = - \frac{1}{N} \; \frac{Ng}{1+Ng}
\; \; \frac{\sqrt{s_i s_l}}{s_l+s_i}.
\end{eqnarray}
Which is the final form of the solution for the particle-hole continuum.
Observe that ${\cal M}_{il}$ has a diagonal term of order zero in $N^{-1}$.
If we have take into account the fact that $s_l=s_{-l}$ it is easy
to show that this term has the form ${\cal M}^0_{il} = \frac{1}{\sqrt{2}}
\left(\delta_{i,l} + sgn(i-l) \delta_{i,-l}\right)$. This term corresponds
to a rotation of the identity matrix in the bosonic representation. Any
calculation of physical quantities with these matrix would give
the same result as for the
non-interacting system. In terms of the Landau theory the rotation of
the bosonic eigenfunctions is essentially the Landau's adiabatic principle
which states that the states of the interacting system and the non-interacting
system can be linked adiabaticlly \cite{Baym,pines}. The next term in
${\cal M}_{il}$ is order $N^{-1}$ and represents the dressing of the
particle by the interaction, that is, the formation of the polarization
cloud. Although this is an effect of order $N^{-1}$ in the matrix
${\cal M}_{il}$ it may be appreciable, since the number of points of the
Fermi surface is $N$. We will show that this is the case for the
calculation of the quasiparticle residue. Furthermore, observe that
the matrix ${\cal N}_{il}$ has contributions of order $N^{-1}$ only
and therefore this matrix is null for the non-interacting system in
the thermodynamic limit. Moreover,
the matrix ${\cal N}_{il}$ carries no resonant terms in the denominator
of (7.24). This can be understood easily from the
discussion of the preceeding section. From definition (7.7) we observe
that while ${\cal M}_{il}$ links points on the same hemisphere of the
Fermi surface the matrix ${\cal N}_{il}$ links points in opposite
hemispheres. What we are showing here is that, since we choose a regular
interaction, in the thermodynamic
limit the contribution from opposite points of the Fermi surface
vanishes completely. All the physics of the
Fermi liquids is defined in just one hemisphere of the Fermi surface.
We will show in the last section of this article that in one dimension
this is not the case because the Fermi surface has a finite number of
points.

Another interesting feature of the solution (7.24) is its dependence on
the interaction. Observe that in terms of our original variables the
coefficient of the correction $N^{-1}$ in the matrices can be written
as $\frac{N(0) U(q)}{1 +N(0) U(q)}$. Thus, if we have a long range
interaction, a Coulomb interaction for instance, this term is
{\it always} finite and we can replace it for a effective interaction
which is local at long wavelengths. In the next section we show that this
term has close relationship with the quasiparticle residue.
Moreover, as we will show
in section IX this term represents the {\it screening} of long
range interactions which arises naturally in the bosonic formalism.

\section{The fermion propagator for a Fermi liquid in two dimensions}

The fermion propagator can be evaluated by using equation (3.3) and
the Bogoliubov transformation (7.7). It is a tedious, but trivial
algebra to show that the fermion propagator for two points, $i$ and $j$,
at the same hemisphere of the Fermi surface can be written as,
\begin{eqnarray}
\langle i, \vec{r}, t| j,0\rangle = \exp\left\{ \sum_{\vec{q},l}
\frac{1}{2 N(0) V q v_F} \left(\frac{{\cal M}_{il} {\cal M}_{il}
+ {\cal N}_{il} {\cal N}_{il}}{s_i} + \frac{{\cal M}_{jl} {\cal M}_{jl}
+ {\cal N}_{jl} {\cal N}_{jl}}{s_j} - \right.\right.
\nonumber
\\
\left. \left. - 2 \frac{{\cal M}_{il} {\cal M}_{jl} e^{i\left(q v_F S_l t -
\vec{q} \cdot \vec{r}\right)} + {\cal N}_{il} {\cal N}_{jl}
e^{i\left(q v_F S_l t + \vec{q} \cdot \vec{r}\right)}}
{\sqrt{s_i s_j}}\right)\right\},
\end{eqnarray}
while for two points in opposite sides of the Fermi surface we find,
\begin{eqnarray}
\langle i, \vec{r}, t| j,0\rangle = \exp\left\{ \sum_{\vec{q},l}
\frac{1}{2 N(0) V q v_F} \left(\frac{{\cal M}_{il} {\cal M}_{il}
+ {\cal N}_{il} {\cal N}_{il}}{s_i} + \frac{{\cal M}_{jl} {\cal M}_{jl}
+ {\cal N}_{jl} {\cal N}_{jl}}{s_j} - \right.\right.
\nonumber
\\
\left. \left. - 2 \frac{{\cal M}_{il} {\cal N}_{jl} e^{i\left(q v_F S_l t -
\vec{q} \cdot \vec{r}\right)} + {\cal N}_{il} {\cal M}_{jl}
e^{i\left(q v_F S_l t + \vec{q} \cdot \vec{r}\right)}}
{\sqrt{s_i s_j}} \right)\right\}.
\end{eqnarray}

Therefore our task is to calculate the sums in these two expressions.
If we use the result (7.7) we find,
\begin{eqnarray}
\sum_l {\cal M}_{il} {\cal M}_{jl} e^{i v_F q S_l t} =
\delta_{i,j} e^{i v_F q s_i t} + \frac{g}{1+Ng}
\frac{\sqrt{s_i s_j}}{s_i-s_j} \; \; \left(
e^{i s_i v_F q t} - e^{i s_j v_F q t}\right) +
\nonumber
\\
+ \left(\frac{g}{1+Ng}\right)^2 \sqrt{s_i s_j}
\sum_l \frac{s_l e^{i v_F q s_l t}}
{\left(s_l-s_i\right)\left(s_l-s_j\right)}.
\end{eqnarray}
Notice that the last term in the above expression has a double
singularity at $s_l=s_i$ and $s_l=s_j$. We
can extract this singularity using the Poincare's theorem,
\begin{equation}
{\cal P} \frac{1}{\left(s_l-s_i\right)\left(
s_l-s_j\right)} =
{\cal P} \frac{1}{s_i-s_j} \left(\frac{1}{s_l-s_i} -
\frac{1}{s_l-s_j}\right) + \pi^2 \delta\left(s_l-s_i\right)
\delta\left(s_l-s_j\right).
\end{equation}
The double delta function can be rewritten in terms of the angles at
the Fermi surface due to the definitions in (7.6). Indeed, it is trivial
to show that,
\begin{equation}
\delta\left(s_l-s_i\right) = \delta\left(\cos \left(\theta_l\right)
- \cos \left(\theta_i\right) \right) = \frac{N}{2 \pi} \frac{\delta_{i,l}}
{\left| \sin\left(\theta_l\right)\right|} = \frac{N}{2 \pi} \frac{\delta_{i,l}}
{\sqrt{1-s_l^2}},
\end{equation}
where we have used $\theta_l = \frac{2 \pi l}{N}$. If we substitute (8.4)
in (8.3) we finally find,
\begin{eqnarray}
\sum_l {\cal M}_{il} {\cal M}_{jl} e^{i v_F q S_l t} =
\left(1+\frac{1}{4} \left(\frac{g N}{1+Ng}\right)^2
\frac{s_i^2}{1-s_i^2} \right) \delta_{i,j} e^{i v_F q s_i t}
+ {\cal O}\left(\frac{1}{N}\right),
\end{eqnarray}
where we have included only the terms of order zero in $N^{-1}$, that is,
the only terms that survive in the thermodynamic limit.
By the same token we can show that,
\begin{eqnarray}
\sum_l {\cal N}_{il} {\cal N}_{jl} e^{i v_F q S_l t} =
{\cal O}\left(\frac{1}{N}\right).
\end{eqnarray}
As we have explained before this result shows that processes which
involve particle-hole pairs at opposite sides of the Fermi surface
disappear in the thermodynamic limit.

If we substitute (8.6) and (8.7) in the expression for the Green's function
(8.1) we end up with the following expression,
\begin{equation}
\langle i, \vec{r}, t| j,0\rangle = \langle i, \vec{r}, t| j,0\rangle_0 \;
\exp\left\{ - \sum_{\vec{q}} \frac{1}{ 4 N(0) v_F V}
\left(\frac{N(0) U(q)}{1+N(0) U(q)}\right)^2 \frac{q_N}{q_T^2}
\left(1 - e^{i \left(v_F q_N t - \vec{q} \cdot \vec{r}\right)} \right)
\right\},
\end{equation}
where $\langle i, \vec{r}, t| j,0\rangle_0 $ is the propagator for the
non-interacting system calculated in (5.15). Observe again the form of
the interaction coefficient which appears in the {\it screening} form as
we have pointed out in previous sections. That is, long range potentials
will not change the quasiparticle residue because, at long wavelengths the
potential seen by the fermions is purely local. In this paper, in order to
simplify the calculation we just assume a local interactions from the
beggining, $U(q)=U$, however we must keep in mind that screening is
a natural consequence of bosonization in dimensions higher than one.

Applying the same methods we
have used in section V we make the replacement,
\begin{equation}
\int \frac{d q_T}{q_T^2}
\rightarrow \frac{N(0) v_F}{\Lambda^2},
\end{equation}
and we are left with the integral,
\begin{equation}
\int_{0}^{\infty} dq_N \; q_N \; e^{- \frac{q_N}{\Lambda}}
\left(1 - e^{i q_N \left(v_F t - \vec{n}_j \cdot \vec{r}\right)} \right)
= \Lambda^2 - \frac{\Lambda^2}{\left(1 - i \Lambda \left(v_F t - \vec{n}_j
\cdot \vec{r}\right)\right)^2} \approx \Lambda^2,
\end{equation}
where we have disregarded the last term in (8.10) since we are looking
at the asymptotic state of the system.

Substituting (8.10) and (8.9) in (8.8) we find,
\begin{equation}
\langle i, \vec{r}, t| j,0\rangle = Z_F \langle i, \vec{r}, t| j,0\rangle_0
\end{equation}
where,
\begin{equation}
Z_F = \exp\left\{-\frac{1}{4} \left(\frac{N(0)U}{1+N(0) U}\right)^2
\right\},
\end{equation}
is the quasiparticle residue. Observe that this form of the quasiparticle
residue has clear non-perturbative character. In the limit of weak coupling,
that is, $ N(0)U \to 0$, we find,
\begin{equation}
Z_F \approx 1 - \frac{(N(0)U)^2}{4}
\end{equation}
which is the expected form of the RPA solution for local interactions
\cite{kadanoff,gordon}. In the strong coupling limit, $ N(0)U \to \infty$,
it is easy to show that,
\begin{equation}
Z_F \approx 0.78 \left(1 + \frac{1}{2 N(0)U}\right),
\end{equation}
which, once more, has a non-perturbative character. Observe that
the quasiparticle residue never vanishes and it actually saturates.
That is, the Fermi liquid state survives for any strength of the
interaction. Also observe that the propagator for opposite hemispheres
of the Fermis surface, (8.2), vanishes in the thermodynamic limit.

We want to emphasize that
the result (8.14) is not completely universal. The numerical factor in
front of equation (8.14) can change due to geometric factors. We have used
Haldane's
construction \cite{haldane} of a sphere with radius $\Lambda$ at each
point of the Fermi surface. In this construction the Fermi surface is
locally flat. If we use another different construction, such as in
ref.\cite{houghton}, we would get a different numerical factor. This factor
is irrelevant and lead to a small renormalization of the quasiparticle
residue. But this is not the important point here, the most striking
result is the survivor of the Fermi liquid state for any strength of
the interaction potential in the case of isotropic, local, interactions.

\section{Screening of a scalar potential}
In the previous section we have shown that screening appears naturally
in the exact solution of the Landau fixed point if we use bosonization.
In this section we reconsider the problem of screening of external
probes by the interacting Fermi system in terms of the bosonized theory.
This is a well understood problem in the framework of Fermi liquid
theory and the results of our bosonized theory naturally agree with the
Fermi liquid theory results.

In this section we examine the behavior
of the system under a scalar external field whose external hamiltonian
is written as,
\begin{equation}
H_{ext} = \frac{1}{V} \sum_{\vec{q}} V_{ext}(q,t) \rho(\vec{q})
\end{equation}
where $\rho(\vec{q})$ is the Fourier component of the charge
density of the system which is written in terms of the operators
(2.1) as,
\begin{equation}
\rho(\vec{q}) = \sum_{\vec{k}}n_{\vec{q}}(\vec{k}).
\end{equation}

Therefore, we consider external hamiltonians of the form,
\begin{equation}
H_{ext}=\frac{1}{V} \sum_{\vec{q},\vec{k}} V_{ext}(q,t) n_{\vec{q}}(\vec{k})
\end{equation}
which, in terms of bosons, is written as,
\begin{equation}
H_{ext} = \sum_{\vec{q},\vec{k}_F,\vec{v}_{\vec{k}_F} \cdot \vec{q}>0}
\left(\frac{
N_{\Lambda}(\vec{k}_F)|\vec{q} \cdot \vec{v}_{\vec{k}_F}|}{V}\right)^{1/2}
V_{ext}(q,t) \left(b^{\dag}_{\vec{q}}(\vec{k}_F)+b_{\vec{q}}(\vec{k}_F)\right).
\end{equation}

It will be useful later to observe that the electronic density is given
in terms of the bosons as,
\begin{equation}
\rho(\vec{q}) = \sum_{\vec{k}_F,\vec{v}_{\vec{k}_F} \cdot \vec{q}>0} \left(
N_{\Lambda}(\vec{k}_F) V |\vec{q} \cdot \vec{v}_{\vec{k}_F}|\right)^{1/2}
\left(b^{\dag}_{\vec{q}}(-\vec{k}_F)+b_{\vec{q}}(\vec{k}_F)\right).
\end{equation}

Here we consider the effect of the potential (9.4) in
a Fermi liquid in order to see what kind of effect it can cause.
The hamiltonian of the system in this case is given by $H=H_{FL} +
H_{ext}$ where $H_{FL}$ is given in (7.1).

The equation of motion for the fields is given by the saddle point
equation for the action in (3.8),
\begin{eqnarray}
i\frac{\partial \phi_{\vec{q}}(\vec{k}_F)}{\partial t}=
-\vec{q} \cdot \vec{v}_{\vec{k}_F} \phi_{\vec{q}}(\vec{k}_F)
-U(q) \sum_{\vec{k'}_F} \left(|\vec{q} \cdot \vec{v}_{\vec{k}_F}|
|\vec{q} \cdot \vec{v}_{\vec{k'}_F}| N_{\Lambda}(\vec{k}_F)
N_{\Lambda}(\vec{k'}_F)\right)^{1/2}
\nonumber
\\
\left( \phi_{\vec{q}}(\vec{k'}_F) + \phi^{*}_{\vec{q}}(-\vec{k'}_F)\right)
-\left(\frac{|\vec{q} \cdot \vec{v}_{\vec{k}_F}| N_{\Lambda}(\vec{k}_F)}{V}
\right)^{1/2} V_{ext}(q,t)
\end{eqnarray}
which, after a Fourier transform, gives,
\begin{eqnarray}
\left(|\vec{q} \cdot \vec{v}_{\vec{k}_F}| N_{\Lambda}(\vec{k}_F) V\right)^{1/2}
\phi_{\vec{q}}(\vec{k}_F,\omega) =
\frac{\vec{q} \cdot \vec{v}_{\vec{k}_F}}{\omega - \vec{q} \cdot
\vec{v}_{\vec{k}_F}} N_{\Lambda}(\vec{k}_F) \Biggl\{
V_{ext}(q,\omega) \Biggr.
\nonumber
\\
\left. +U(q) \sum_{\vec{k'}_F} \left(|\vec{q} \cdot
\vec{v}_{\vec{k'}_F}| N_{\Lambda}(\vec{k'}_F) V \right)^{1/2}
\left(\phi_{\vec{q}}(\vec{k'}_F,\omega)+
\phi^{*}_{\vec{q}}(-\vec{k'}_F,\omega)\right)\right\}.
\end{eqnarray}

Summing over $\vec{k}_F$ on both sides of (9.7) and using (9.5)
we find,
\begin{equation}
\langle \rho(\vec{q},\omega) \rangle = \Pi(\vec{q},\omega)
\left(V_{ext}(q,\omega) + U(q) \langle \rho(\vec{q},\omega) \rangle\right)
\end{equation}
where,
\begin{equation}
\Pi(\vec{q},\omega) = \sum_{\vec{k}_F} N_{\Lambda}(\vec{k}_F)
\frac{\vec{q} \cdot \vec{v}_{\vec{k}_F}}{\omega -
\vec{q} \cdot\vec{v}_{\vec{k}_F}}
\end{equation}
is the polarization function we have found in (7.18).

The interpretation for (9.8) is very simple: the external
potential produces a polarization of the fermionic gas which
shields the interaction at long distances. Therefore, instead of
the bare external potential $V_{ext}(q,\omega)$ a new polarization
potential appears, $U(q) \langle \rho(\vec{q},\omega) \rangle$, and
the effective potential which is felt by
the fermions is given by,
\begin{equation}
U_{eff}(\vec{q},\omega) = V_{ext}(q,\omega) +
U(q)\langle \rho(\vec{q},\omega)\rangle.
\end{equation}

The effective potential can be obtained from the bare potential
using (9.8) and (9.10). Solving (9.8) for the density one gets,
\begin{equation}
\langle \rho(\vec{q},\omega) \rangle = \frac{\Pi(\vec{q},\omega)
V_{ext}(q,\omega)}{1- U(q) \Pi(\vec{q},\omega)}.
\end{equation}
Substituting (9.11) in (9.10) we finally find,
\begin{equation}
U_{eff}(\vec{q},\omega) = \frac{V_{ext}(q,\omega)}{\epsilon(\vec{q},\omega)}
\end{equation}
where $\epsilon(\vec{q},\omega)$ is the dielectric function of the
fermionic system which is given by,
\begin{equation}
\epsilon(\vec{q},\omega)= 1 - U(q) \Pi(\vec{q},\omega).
\end{equation}
The expressions (9.12) and (9.13)
are the RPA results for the electronic gas
\cite{Baym,pines,abrikosov,nozieres,kadanoff}. Since we are dealing
with the {\it exact} diagonalization of the fermionic system at
{\it long} wavelengths, it is natural to recover the RPA approximation
as the exact result in this limit since, as it is well known, its long
wavelength limit saturates the sum rules.

In the limit of interest, namely, low frequency and small momenta,
the polarization function can be easily obtained. From (7.18) we have,
\begin{equation}
\Pi(\vec{q},\omega) = \frac{N(0)}{S_d} \int d\Omega \frac{\cos(\theta)}
{\frac{\omega}{v_F q} - \cos(\theta) + i\eta}
\end{equation}
where we have included a small imaginary part for the frequency,
$\Omega$ is the solid angle and $\theta$ is the angle between the
Fermi velocity and $\vec{q}$. Observe that the polarization function
is only a function of $\frac{\omega}{v_F q}$. We can also rewrite (9.14) as,
\begin{equation}
\Pi(\vec{q},\omega) = N(0)\left({\cal P}\int \frac{d\Omega}{S_d}
\frac{\cos(\theta)}{\frac{\omega}{v_F q} - \cos(\theta)}
-i\pi \int \frac{d\Omega}{S_d} \cos(\theta) \delta\left(\frac{\omega}{v_F q}
- \cos(\theta)\right)\right)
\end{equation}
where ${\cal P}$ means principal value.

In the limit of $\frac{\omega}{v_F q}<<1$ we obtain
\begin{equation}
\Pi(\vec{q},\omega) = - N(0) \left(1 + i \pi \gamma_d
\frac{\omega}{v_F q}\right)
+{\cal O}\left(\frac{\omega^2}{v^2_F q^2}\right),
\end{equation}
where $\gamma_d$ is a numerical factor which depends on the spatial
dimensionality, $d$, of the system ($\gamma_1=0,\gamma_2=1/\pi,
\gamma_3=1/2$).
The second term in r.h.s. of (9.16) is the Landau damping
term due to the decay of
the bosonic modes into the particle-hole continuum. Observe that
in one dimensional systems the bosonic modes never decay
due fact that $\gamma_1=0$. It means that the true excitations of
the one dimensional system are collective modes.
In higher dimensions this is not true because there is always
a particle hole continuum. In section XI we will see that this
result has important consequences for the calculation of the Green's
function.

At this point it is tempting to follow the conventional approximation of
Fermi liquid theory and to argue that the fermion-fermion interaction
should also get screened just as much as an external probe is. However
this is a delicate argument which is only justified by its success in
explaining experiments. One should keep in mind that this is essentially
a perturbative argument, motivated by a partial resummation of the
perturbation theory series (namely the sum of all the bubble diagrams or RPA).
In earlier sections of this paper, we showed
that since the tangential fluctuations do not
change the energy, they mainly give rise to a density of states. In the
conventional ``screening first" argument of Fermi liquid theory (in the
RPA approximation\cite{abrikosov}) one finds an effective electron-electron
interaction
of the form
\begin{equation}
U_{eff}(\vec{q},\omega) = \frac{1}{U^{-1}(q) - \Pi(\vec{q},\omega)}.
\end{equation}
If we substitute the asymptotic form of the polarization $\Pi$ the RPA
expression for the effective potential becomes,
\begin{equation}
U_{eff}(\vec{q},\omega) \approx \frac{U(q)}{1+ U(q) N(0)}.
\end{equation}
This expression shows that, within RPA, screening is an effect caused by
the density of states of the Fermi system. Notice, however, that this is
exactly the same expression that appears in equation (8.8) when we
diagonalized exactly the bosonic problem. Thus, bosonization in dimensions
higher than one produces screening naturally and we do not have to put it
{\it by hand} as in the usual approach in condensed matter physics.
Furthermore,
notice that our arguments in this section do not depend on the dimensionality
of the system, therefore, we conclude that in any number of dimensions the
screening of an external potential has RPA character. However, as we will
show at the last section of this paper, in one dimension the fermion-fermion
interaction is {\it not} screened. In particular, this result leads to the
vanishing of the Green's function for the Coulomb potential in one dimension.
Once more, if we have screened the potential {\it by hand} we would get
a wrong result.

\section{Coupling to Gauge fields}

In the last two sections we showed how screening arises in bosonization of
fermionic
systems with long range scalar interactions. The reason for that is
that the fermionic system resembles a liquid which can sustain
longitudinal oscillations. However, if transverse
oscillations are present the physics of the system changes completely.

Let us consider a system of fermions in $D>1$ space dimensions. In one
space dimension, all gauge fields are purely longitudinal and, hence,
the
interactions they mediate are equivalent to Coulomb-like interactions of
the form discussed in the previous section. Suppose, exactly as in the
previous section, that we
couple the fermionic system via minimal coupling with an external vector
field. The hamiltonian can be written as \cite{reizer},
\begin{equation}
H = H_0 + H_{F-G}
\end{equation}
$H_0$ is the non-interacting hamiltonian and
\begin{equation}
H_{F-G} = g \sum_{\vec{q},\vec{k}} \vec{v}_{\vec{k}-\frac{\vec{q}}{2}}
\cdot \vec{A}_{-\vec{q}} n_{\vec{q}}(\vec{k})
+ \frac{g^2}{2 m^{*}} \sum_{\vec{q},\vec{q'},\vec{k}}
\vec{A}_{\vec{q}} \cdot \vec{A}_{\vec{q'}} n_{\vec{q}+\vec{q'}}(\vec{k})
+H_G
\end{equation}
is the fermion-gauge part of the hamiltonian where $m^*$ is the
mass of the fermions. The first and the second terms on the r.h.s.
of (10.2) give the coupling between the gauge fields and the fermionic
system ($g$ is the coupling constant of the theory). The last term on
the r.h.s of (10.2) is a pure gauge field term usually describing the energy of
the gauge field or possible external current terms which generate the
field in the system.

Since we are dealing with small momentum transfer only, we notice that
the dominant contribution from the second term comes from $\vec{q}=-\vec{q'}$
(the terms with $\vec{q}=0$ or $\vec{q'}=0$ are not present in order to
preserve charge neutrality in the system).
Using $\sum_{\vec{k}}n_0(\vec{k})=N_f$ where $N_f$ is the
number of fermions we can rewrite (10.2) in terms of the bosons operators
approximately as,
\begin{eqnarray}
H_{F-G} \approx g \sum_{\vec{q},\vec{k}_F,\vec{v}_{\vec{k}_F} \cdot \vec{q}>0}
\left(|\vec{q} \cdot \vec{v}_{\vec{k}_F}| N_{\Lambda}(\vec{k}_F) V\right)^{1/2}
\vec{v}_{\vec{k}_F-\frac{\vec{q}}{2}}
\cdot \left(\vec{A}_{-\vec{q}} \, b^{\dag}_{\vec{q}}(\vec{k}_F)
+\vec{A}_{\vec{q}} \, b_{\vec{q}}(\vec{k}_F)\right)
\nonumber
\\
+ \frac{g^2 N_f}{2 m^{*}} \sum_{\vec{q}} |\vec{A}_{\vec{q}}|^2
+H_G
\end{eqnarray}

The generating functional is obtained exactly as before and the
equations of motion for the bosonic fields are simply,
\begin{equation}
i\frac{\partial \phi_{\vec{q}}(\vec{k}_F)}{\partial t}=
-\vec{q} \cdot \vec{v}_{\vec{k}_F} \phi_{\vec{q}}(\vec{k}_F)
- g \left(|\vec{q} \cdot \vec{v}_{\vec{k}_F}|
N_{\Lambda}(\vec{k}_F) V \right)^{1/2} \vec{v}_{\vec{k}_F-\frac{\vec{q}}{2}}
\cdot \vec{A}_{\vec{q}}.
\end{equation}
And for the gauge fields we find,
\begin{eqnarray}
\sum_j \left[D_0^{-1}\left(\frac{\partial}{\partial t},\vec{q}\right)
\right]_{i,j}
\left[\vec{A}_{\vec{q}}(t)\right]_j=
g \sum_{\vec{k}_F} \left[\vec{v}_{\vec{k}_F-\frac{\vec{q}}{2}}\right]_i
\left(\frac{|\vec{q} \cdot \vec{v}_{\vec{k}_F}|
N_{\Lambda}(\vec{k}_F)}{V}\right)^{1/2} \phi_{\vec{q}}(\vec{k}_F,t)
\nonumber
\\
+ \frac{g^2 N_f}{m^{*} V} \left[\vec{A}_{\vec{q}}(t)\right]_i
+\left[\vec{J}_{ext}(\vec{r},t)\right]_i
\end{eqnarray}
where $D_0^{-1}\left(\frac{\partial}{\partial t},\vec{q}\right)$ is the
bare propagator for the gauge field (which in general is a tensor with
spatial components $i,j=1,2,\cdots d$ where $d$ is the number of spatial
dimensions) and $\vec{J}_{ext}(\vec{r},t)$ is some external current. Eqs.(10.4)
and (10.5) are coupled. They can be solved by a Fourier transform.
Solving for (10.4) we find,
\begin{equation}
\phi_{\vec{q}}(\vec{k}_F,\omega) = g \frac{\left(|\vec{q} \cdot
\vec{v}_{\vec{k}_F}|N_{\Lambda}(\vec{k}_F) V \right)^{1/2}}{\omega-\vec{q}
\cdot \vec{v}_{\vec{k}_F}} \vec{v}_{\vec{k}_F-\frac{\vec{q}}{2}}
\cdot \vec{A}_{\vec{q}}(\omega).
\end{equation}
Substituting this result in (10.5) one finds,
\begin{eqnarray}
\sum_j \left[D_0^{-1}\left(\omega,\vec{q}\right)\right]_{i,j}
\left[\vec{A}_{\vec{q}}(\omega)\right]_j=
g^2 \sum_{\vec{k}_F,l} N_{\Lambda}(\vec{k}_F)
\frac{\vec{q} \cdot \vec{v}_{\vec{k}_F}}{\omega -
\vec{q} \cdot\vec{v}_{\vec{k}_F}}
\left[\vec{v}_{\vec{k}_F-\frac{\vec{q}}{2}}\right]_i
\left[\vec{v}_{\vec{k}_F-\frac{\vec{q}}{2}}\right]_l
\left[\vec{A}_{\vec{q}}(\omega)\right]_l
\nonumber
\\
+ \frac{g^2 n_f}{m^{*}} \left[\vec{A}_{\vec{q}}(\omega)\right]_i
+\left[\vec{J}_{ext}(\vec{q},\omega)\right]_i
\end{eqnarray}
where $n_f = N_f/V$ is the density of fermions. Eq.(10.7) can
be rewritten in a more appealing form as,
\begin{equation}
\sum_j \left[D^{-1}\left(\omega,\vec{q}\right)\right]_{i,j}
\left[\vec{A}_{\vec{q}}(\omega)\right]_j=\left[\vec{J}_{ext}
(\vec{q},\omega)\right]_i
\end{equation}
where,
\begin{equation}
\left[D^{-1}\left(\omega,\vec{q}\right)\right]_{i,j}=
\left[D_0^{-1}\left(\omega,\vec{q}\right)\right]_{i,j}
- g^2 \left( N(0) v^2_F \delta_{i,j} +
\sum_{\vec{k}_F} N_{\Lambda}(\vec{k}_F)
\frac{\vec{q} \cdot \vec{v}_{\vec{k}_F}}{\omega -
\vec{q} \cdot\vec{v}_{\vec{k}_F}}
\left[\vec{v}_{\vec{k}_F-\frac{\vec{q}}{2}}\right]_i
\left[\vec{v}_{\vec{k}_F-\frac{\vec{q}}{2}}\right]_j
\right).
\end{equation}
is the effective propagator for the gauge fields \cite{pines}.
We have used that $m^{*} = n_F/(v_F^2 N(0))$.
In general it is very difficult to calculate the correction
to the bare propagator. However, for an isotropic system, due
to the symmetry, only the diagonal terms survive. For long
wavelengths ($q<<k_F$) and small frequencies ($\omega<< v_F k_F$)
one finds,
\begin{equation}
\left[D^{-1}\left(\omega,\vec{q}\right)\right]_{i,j} =
\left[D_0^{-1}\left(\omega,\vec{q}\right)\right]_{i,j} -
v^2_F g^2 \left(N(0)+ \Pi(\vec{q},\omega)\right)\delta_{i,j}
\end{equation}
where $\Pi(\vec{q},\omega)$ is defined in (7.18).

In the limit of interest, namely $\frac{\omega}{v_F q}<<1$, we can use
the result (10.16) and rewrite the renormalized propagator as,
\begin{equation}
\left[D^{-1}\left(\omega,\vec{q}\right)\right]_{i,j} =
\left[D_0^{-1}\left(\omega,\vec{q}\right)\right]_{i,j} +
i \omega^2_{pG} \frac{\omega}{v_F q} \delta_{i,j}
\end{equation}
where
\begin{equation}
\omega^2_{pG} = \frac{\pi g^2 n_f \gamma_d}{m^{*}} = \pi \gamma_d
g^2 N(0) v_F^2
\end{equation}
is the plasma frequency associated with the oscillations of the
electronic system due to the coupling to gauge field.

Observe that, contrary to the scalar case of section IX, the
zero frequency gauge fields are not affected by fluctuations of the
Fermi system, that is,
\begin{equation}
\left[D\left(0,\vec{q}\right)\right]_{i,j} =
\left[D_0\left(0,\vec{q}\right)\right]_{i,j}.
\end{equation}
The cancellation of the density of states term in eq.~(10.11) implies
that the only effect of
electron correlations at low frequencies and small wavevectors is a
{\it damping} (not screening) of the transverse gauge
fields. This is the well known phenomenon of Landau damping of
transverse gauge fields in metals.
A non cancelling density of states would imply a gap in the spectrum of
fluctuations of the transverse gauge fields and the expulsion of static
gauge fields, namely, a Meissner effect. This is what happens in a
superconducting state.

Let us assume, for the sake of the argument, that the bare propagator
has the following form,
\begin{equation}
\left[D_0\left(\omega,\vec{q}\right)\right]_{i,j} =
\frac{1}{\omega^2 - v_G^2 q^2}
\left[{\bar D}_0\left(\omega, {\vec q}\right)\right]_{i,j}
\end{equation}
where $v_G$ is the velocity of propagation of the gauge fields.
$\left[{\bar D}_0\left(\omega, {\vec q}\right)\right]_{i,j}$ a tensor whose
form depends on the choice of gauge and it has an analytic dependence in
$\omega$ and $\vec q$. The renormalized propagator has the
form (up to analytic structure tensors)
\begin{equation}
D\left(\omega,\vec{q}\right) =\frac{1}{\omega^2 - v_G^2 q^2 +
 i \omega^2_{pG} \frac{\omega}{v_F q}}.
\end{equation}
The Landau damping introduces a new physical scale in
the problem. Observe that the interaction
is screened with a frequency dependent screening length of order,
\begin{equation}
l_s(\omega) \sim \left(\frac{v^2_G v_F}{\omega^2_{pG}}\right)^{1/3}
\omega^{-1/3},
\end{equation}
which defines a new scale in the problem.
In the strong coupling limit,
that is, $ g \to \infty$, the plasma frequency is much
larger than the characteristic frequencies of the system ($\omega_{pG}
>>\omega$). In this limit
the characteristic momentum of the system will be cut-off by the
plasma frequency, $\frac{\omega_{pG}}{\sqrt{2 v_F v_G}} > q$ and the
asymptotic form of the propagator is dominated by the Landau term,
namely,
\begin{equation}
D\left(\omega,\vec{q}\right) =\frac{-1}{v_G^2 q^2 - i \omega^2_{pG}
\, \frac{\omega}{v_F q}}.
\end{equation}
In real space and time this form of the propagator implies that the
gauge fields behaves {\it diffusively}.
The same type of propagator is found in the RPA approach \cite{pines,reizer}.

\section{The one dimensional case}

In the previous sections we have shown that in terms of response
functions bosonization gives the same result of the RPA approximation.
This is due to the fact that RPA fulfill all the sum rules at
long wavelengths and low energies, which is the exactly the limit
where bosonization can be applied. Moreover, RPA is valid for high
densities which in our language means that $k_F$ is large compared
with the fluctuations in the system. This would lead us to conclude,
erroneously, that RPA and bosonization are one and the same thing.
Actually, the RPA results are {\it expected} from the bosonization
point of view. We have already shown that bosonization gives rise to screening
naturally, something that you have to put by hand in the RPA approach.
Moreover, RPA is only valid in the weak coupling limit, $N(0)U <<1$
\cite{pines}, and, as we have shown before, this is not the case of
bosonization. It is indeed well known that in one dimension RPA
works fine for correlations functions while it cannot explain the
absence of isolated singularities in the Green's function. In this
section we try to explain the reason for this behavior by comparison
with our results in higher dimensions.

In one dimension all the calculations simplify enormously. The matrices
${\cal M}_{il}$ and ${\cal N}_{il}$ reduce to numbers. The condition (7.8)
can be rewritten in terms of a variable $\zeta (q)$ such that,
\begin{eqnarray}
{\cal M}(q) = \cosh(\zeta (q))
\nonumber
\\
{\cal N}(q)= \sinh(\zeta (q)).
\end{eqnarray}
The other condition in (7.8) is automatically fulfilled. In the one dimensional
case there is no particle hole continuum since the Fermi surface
reduces to two points, that is, $s_1 =1$. The collective mode
equation (7.16) defines the eigenvalue for the collective mode,
\begin{eqnarray}
g(q) \frac{2}{S_0^2(q) -1} = 1
\end{eqnarray}
which is easily solved as,
\begin{equation}
S_0(q) = \sqrt{ 1 + 2 g(q)} = \sqrt{1 + N(0) U(q)}
\end{equation}
where we used (7.2) with $N=2$. The last expression can be put in
a more standard form if we use the notation of section VII and
rewrite the frequency of oscillation of the collective mode as,
\begin{equation}
E_q = v_F q \sqrt{1 + N(0) U(q)}
\end{equation}
which is the well known result for one-dimensional systems \cite{emery}.
The variable $\zeta (q)$ is defined by the solution (7.15), that is,
\begin{equation}
\tanh(\zeta (q)) = \frac{N(q)}{M(q)} = - \frac{S_0-1}{S_0+1} =
- \frac{\sqrt{ 1 + 2 g(q)}-1}{\sqrt{ 1 + 2 g(q)}+1}
\end{equation}
which can be rewritten in a more standard form,
\begin{equation}
\tanh( 2 \zeta (q)) = \frac{g(q)}{1+g(q)}
\end{equation}
which is the expected result \cite{emery}.

Observe that ${\cal N}$ is finite, contrary to higher dimensions.
This is a result of the finite number of Fermi points. We will
now see that this has deep consequences for the one particle
propagator. Indeed, using equation (11.1) we find,
\begin{equation}
\langle i, x, t| j,0\rangle = \langle i, x, t| j,0\rangle_0
\exp\left\{\sum_q \frac{\sinh^2(\zeta (q))}{V N(0) q v_F}
\left(1 - \cos(q x) e^{i q v_F S_0(q) t} \right) \right\},
\end{equation}
at the same Fermi point (right movers, for instance) and,
\begin{equation}
\langle i, x, t| j,0\rangle = 0
\end{equation}
for opposite Fermi points.  This is the well known result for the one
dimensional Luttinger model \cite{emery}.

The first consequence of the finite value of ${\cal N}$ is the
presence of an anomalous dimension. Indeed, suppose the potential
is local. Then the only effect of the interaction on the spectrum
is a renormalization of the Fermi velocity from $v_F$ to
$\tilde{v}_F = v_F \sqrt{1 + N(0)U}$. In this case the integral
(11.7) is easily done (see (5.14)) and the result is,
\begin{equation}
\langle i, x, t| j,0\rangle = \frac{\alpha}{x-\tilde{v}_F t + i\alpha}
\left(\frac{\alpha^2}{x^2 - (\tilde{v}_F t - i\alpha)^2}\right)^{\gamma}
\end{equation}
where
\begin{equation}
\gamma=\frac{1 + N(0)U}{\sqrt{1+N(0)U}} - 1.
\end{equation}
Observe that the above Green's function has an anomalous dimensions
given by $\gamma$ and a branch cut in the spectrum instead of an
isolated singularity. As we discussed before this is a result
of the process (a forward scattering!) that links opposite sides
of the Fermi surface, a process which is suppressed by the presence
of the particle-hole continuum in higher dimensions (in the absence
of nesting, singular interactions or gauge fields).

But this is not the only difference between one and higher dimensions.
In the form of the fermion propagator (11.7) the interactions are not screened.
Suppose, for instance, that we have a Coulomb interaction, that is,
$U(q) = e^2/q^2$. This is the case of the Schwinger model
\cite{schwinger,andersonhiggs} which was studied via bosonization by
Kogut and Susskind \cite{kogut}. In real space the Fourier transform
of the Coulomb potential gives rise to a linear potential and therefore
to confinement. If we calculate the spectrum from (11.4) we get a massive
relativist bosonic theory (the system has a gap at $q=0$). It is easy
to verify that the propagator in (11.7) has an infrared divergence at
$t \neq 0$ and it vanishes in this limit, that is, the fermions decay
and disappears completely from the spectrum, only the collective mode
is left. However, as we have shown
in the section IX the one dimensional fermion gas screens external
probes as expected, that is, in the RPA form. Of course, if we have
screened the potential first, as in the RPA approach, we would never get
this amazing result.

Therefore, the existence of the Luttinger fixed point in one dimension
and the presence of anomalous dimensions
is just a result of the lack of phase space (or due to a finite number
of points in the Fermi surface).

\section{Concluding remarks}

In our previous papers \cite{us} we studied the transport
and thermodynamic properties of Fermi liquids. In this paper
we have studied the Landau fixed point of a Fermi liquid and the
associated one-particle propagator using the method of bosonization
in arbitrary dimensions.

We have shown that it is possible to bosonize a theory of interacting
fermions in the restricted Hilbert space of states close to the Fermi
surface. The bosonization is based on the algebra of the particle-hole
operators in this Hilbert space and in the introduction of bosonic
coherent states which generate deformation of the Fermi surface.
{}From the coherent states it is possible to define a generating functional
which is a path integral over the histories of the Fermi surface.
The fields which propagate on the Fermi surface are sound waves
which can be viewed as coherent superposition of particle-hole pairs.

We have shown that from the construction of the fermion operator
via coherent states we obtain the correct one-particle propagator
which represents a fermion moving with the Fermi velocity.
We also discuss the terms which appear in the interacting hamiltonian
and and their relevance for the fixed point. It is obtained that
processes which involve particle-holes on one hemisphere of the
Fermi surface are associate with the Landau fixed point and the
presence of anomalous dimensions, and the absence of singularities
in the Green's function is due to operators which mix different
sides of the Fermi surface. These operators appear in one dimension
and in systems with nested Fermi surface due to lack of phase space
for scattering or in the presence of singular interactions (such as
in the case of gauge interactions). It is important to stress here
that we are including {\it forward scattering} only and therefore
our results are valid for {\it gaussian fixed points}.

We show that for the simple case of isotropic interactions the
bosonic hamiltonian can be diagonalized by a generalized Bogoliubov
transformation which mixes different points of the Fermi surface.
We obtain two different types of solutions which represent the
particle-hole continuum and collective mode.
We obtain the fermion propagator in the thermodynamic limit for
local interactions. We show that in dimensions higher than one the
fermion propagator has isolated singularities and the only difference
between the non-interacting
system and the interacting one is the presence of the quasiparticle
residue. We evaluate the quasiparticle residue for any strength of the
interaction (and thus showing the non-perturbative character of the
bosonization approach and its difference from the perturbative approaches
based on ressumation of diagrams, such as RPA) and we show that the
quasiparticle residue is always finite, that is, there is no possibility
of breakdown of Fermi liquid theory for local interactions, exactly
as expected. Furthermore, our results agree with the perturbative ones
in the limit of weak coupling.
Moreover, dynamical screening is a {\it natural} result of the bosonization
method in dimensions higher than one. That is, we obtain that
long range interactions are screened and we do not have to assume
it as it is usual in perturbative theories in condensed matter physics.

We also study the problem of the response of the fermionic gas
to scalar external probes and obtain the RPA result for screening.
This confirms that the bosonization is getting the correct physics
at long wavelengths and low energies since the RPA fulfill the sum rules
in this limit. And we stress once more that it does {\it not}
mean that RPA and bosonization are the same thing because bosonization
is a non-perturbative method which is valid for any strength of the
interaction. We also show that when fermions are coupled to gauge
fields the RPA result is also valid, the interactions are not
screened but there is Landau damping (except in one dimension).
We calculate the form of the effective propagator for the gauge
fields and find the expected form for RPA.

Finally we compare the one dimensional problem, related to a Luttinger
fixed point, with the Landau fixed point. We show that due to the
finite number of Fermi points in one dimension the mixing of points
across the Fermi surface exists and leads to the appearance of anomalous
dimensions in the Green's function. In one dimension there is only a
collective excitation while in higher dimensions there is a particle-hole
continuum which shares spectral weight with the collective mode.
We also show that the perturbative
approach of screening the fermion-fermion interaction by hand would
lead to wrong results in one dimension. While the one dimensional
fermionic system screens external probes in the usual RPA form it
does not screen the fermion-fermion interaction. The bosonization method leads
to the vanishing of the Green's function in the case of Coulomb
interactions in one dimension. In higher dimensions, due to screening,
the Coulomb interaction leads to an effective local interaction which
does no harm to the Fermi liquid behavior.

\acknowledgements

We are deeply indebted to Gordon Baym and Shivaji Sondhi for many useful
and illuminating discussions on this subject. A.~H.~C.~N. thanks
Conselho Nacional de Desenvolvimento
Cient\'ifico e Tecnol\'ogico, CNPq (Brazil), for a scholarship.
This work was supported in part by NSF Grant DMR91-22385 at the
University of Illinois at Urbana-Champaign.

\newpage

\begin{figure}
\caption{a)Forward scattering; b)Backward scattering.}
\end{figure}

\begin{figure}
\caption{Processes described by eqs. (6.3) and (6.4);
a) Forward scattering; b)Backward scattering. The dashed line
indicates the evolution of the state.}
\end{figure}

\begin{figure}
\caption{Processes described by eqs. (6.5) and (6.6);
a) Forward scattering; b)Backward scattering. The dashed line
indicates the evolution of the state.}
\end{figure}

\begin{figure}
\caption{Solution of the equation (7.16). The dashed lines represent
the eigenfrequencies of the modes for the non-interacting system,
the circles show the eingenfrequencies for the interacting system.
a) particle-hole continuum; b) collective mode.}
\end{figure}


\newpage

\begin{references}
\bibitem{Baym}
G.~Baym and C.~Pethick, {\it Landau Fermi-Liquid Theory}, (John Wiley,
New York,1991).
\bibitem{pines}
D.~Pines and P.~Nozi\`eres, {\it Theory of Quantum Liquids},
Vol.~I, (Addison-Wesley, Redwood City, 1989).
\bibitem{abrikosov}
A.~A.~Abrikosov, L.~P.~Gorkov and I.~E.~Dzyaloshinski, {\it Methods of
Quantum Field Theory in Statistical Physics}, (Dover, New York, 1975).
\bibitem{nozieres}
P.~Nozi\`eres, {\it Theory of Interacting Fermi Systems}, (Benjamim, New York,
1964).
\bibitem{kadanoff}
L.~Kadanoff and G.~Baym, {\it Quantum Statistical Mechanics}, (Addison
Wesley, Redwood City, 1991).
\bibitem{bcs}
J.~Bardeen, L.~N.~Cooper and J.~R.~Schriffer, Phys.~Rev. {\bf 108},
1175, (1957).
\bibitem{htc}
J.~G.~Bednorz and K.~A.~Mueller, Z.~Phys.B {\bf 64}, 189, (1986).
\bibitem{varma}
C.~M.~Varma, P.~B.~Littlewood, S.~Schmitt-Rink, E.~Abrahams and
A.~E.~Ruckenstein, Phys.~Rev.~Lett {\bf63}, 1996, (1989).
\bibitem{anderson}
P.~W.~Anderson, Phys.~Rev.~Lett. {\bf 64}, 1839, (1990);
{\bf 65}, 2306, (1990); {\bf 66},3226, (1991); {\bf 67}, 2092, (1991);
{\bf 71}, 1220, (1993);
S.~Chakravarty, A.~Sudb\o, P.~W.~Anderson and S.~Strong,
Science {\bf 261}, 337, (1993).
\bibitem{urbana}
D.~Pines, UIUC preprint P-93-06-050 (1993).
\bibitem{bosonization}
S.~Tomonaga, Prog.~Theor.~Phys. {\bf 5}, 544 (1950); E.~Lieb and
D.~C.~Mattis, J.~Math.~Phys. {\bf 6}, 304 (1964); A.~Luther
and I.~Peschel, Phys.~Rev.B {\bf 9}, 2911 (1974); S.~Coleman,
Phys.~Rev.D {\bf 11}, 2088 (1975); S.~Mandelstam, Phys.~Rev.D {\bf 11}, 3026
(1975).
\bibitem{luther}
A.~Luther, Phys.~Rev.B {\bf 19}, 320, (1979).
\bibitem{haldane}
F.~D.~M.~Haldane, private communication and Helv.~Phys.~Acta. {\bf 65},
152 (1992).
\bibitem{houghton}
A.~Houghton and B.~Marston, Phys.Rev.B {\bf 48}, 7790, (1993).
\bibitem{us}
A.~H.~Castro Neto and E.~H.~Fradkin, Phys.Rev.Lett. {\bf 72}, 1393, (1994)
and to appear in Phys.Rev.B {\bf 49}.
\bibitem{lee}
P.~A.~Lee, Phys.~Rev.~Lett. {\bf 63}, 680, (1989); P.~A.~Lee and
N.~Nagaosa, Phys.~Rev.B {\bf 46}, 5621, (1992); G.~Baskaran and
P.~W.~Anderson,  Phys.~Rev.B {\bf 37}, 580, (1988);
L.~B.~Ioffe and A.~I.~Larkin, Phys.~Rev.B {\bf 39},
8988, (1989); B.~Blok and H.~Monien, Phys.~Rev.B {\bf 47}, 3454,
(1993).
\bibitem{ana}
Ana Lopez and Eduardo Fradkin, Phys.~Rev.B {\bf 44}, 5246 (1991).
\bibitem{halperin}
B.~I.~Halperin, P.~A.~Lee and N.~Read, Phys.~Rev.B {\bf 47},
7312, (1993).
\bibitem{houghton2}
A.Houghton, H.-J.Kwon and J.B.Marston, ``On the stability and single-particle
properties of Bosonized Fermi liquids" , cond-mat/9310043; A.Houghton,
H.-J.Kwon, J.B.Marston and R.Shankar, ``Coulomb interaction and the Fermi
liquid state: solution by bosonization", cond-mat/9312067.
\bibitem{next}
A.~H.~Castro Neto and E.~H.~Fradkin, in preparation.
\bibitem{klauder}
J.~Klauder and B.~S.~Skagerstam, {\it Coherent States}, (World Scientific,
singapore, 1985).
\bibitem{halpern}
M.B.Halpern, Phys.Rev.D {\bf 12}, 1684, (1975).
\bibitem{carneiro}
C.~Pethick and G.~Carneiro, Phys.~Rev.A {\bf 7}, 304, (1973).
\bibitem{guinea}
J.~Gal\'an, F.~Guinea, J.~A.~Verg\'es, G.~Chiappe and E.~Louis,
Phys.~Rev.B {\bf 46}, 3136, (1992).
\bibitem{shankar}
R.~Shankar, Physica A {\bf 177}, 530 (1991) and to appear in Rev.~of
Mod.~Phys.~(1993).
\bibitem{ian}
J.R.Engelbrecht and M.Randeria, Phys.Rev.Lett. {\bf 65}, 1032, (1990).
\bibitem{gordon}
We are very grateful to Gordon Baym for pointing out a mistake in the
calculation of the quasiparticle residue in the previous version of
this paper.
\bibitem{reizer}
M.~Yu.~Reizer, Phys.~Rev.B {\bf 39}, 1602, (1989);
Phys.~Rev.B {\bf 40}, 11571, (1989).
\bibitem{emery}
V.~J.~Emery, in {\it Highly Conducting One-Dimensional Solids},Edited
by J.~T.~Devreese, R.~P.~Evrard and V.~E.~van Doren, Plenum Press (1979).
\bibitem{schwinger}
J.~Schwinger, Phys.~Rev. {\bf 125}, 397 (1962)
\bibitem{andersonhiggs}
P.~W.~Anderson, Phys.~Rev. {\bf 130}, 439 (1963).
\bibitem{kogut}
J.Kogut and L.Susskind, Phys.Rev.D {\bf 11}, 3594, (1975).

\end{references}
\end{document}